%% file: main.tex
\documentclass{article}

\usepackage[eandd, preprint]{neurips_2026}

\usepackage[utf8]{inputenc}
\usepackage[T1]{fontenc}
\usepackage{hyperref}
\usepackage{url}
\usepackage{booktabs}
\usepackage{amsfonts}
\usepackage{nicefrac}
\usepackage{microtype}
\usepackage[table]{xcolor}
\usepackage{algorithm}
\usepackage{algpseudocode}
\usepackage{wrapfig}
\usepackage{graphicx}
\usepackage{float}
\usepackage{multirow}
\usepackage{makecell}
\usepackage{amsmath}
\usepackage{pifont}
\usepackage{mdframed}
\usepackage[most]{tcolorbox}
\usepackage{caption}
\usepackage{subcaption}

\definecolor{myMagenta}{RGB}{200, 0, 90}
\definecolor{myGreen}{RGB}{40, 130, 40}
\definecolor{myRed}{RGB}{255, 0, 0}

\hypersetup{
  colorlinks=true,
  linkcolor=myRed,
  citecolor=myGreen,
  urlcolor=myMagenta
}

\newcommand{\cmark}{\textcolor{green!65!black}{\large\ding{51}}}
\newcommand{\xmark}{\textcolor{red}{\large\ding{55}}}
\newcommand{\bi}[1]{\textbf{\textit{#1}}}

\title{LINEAGE: A Benchmark for Empirical Lineage Retrieval in AI Scientific Literature}

\author{Pangjing Wu\thanks{Equal contribution.} \\
  Department of Computing\\
  The Hong Kong Polytechnic University\\
  Hong Kong SAR, China \\
  \And
  Xu Yuan\footnotemark[1] \\
  Department of Computing\\
  The Hong Kong Polytechnic University\\
  Hong Kong SAR, China \\
  \And
  Wenqi Fan \\
  Department of Computing\\
  The Hong Kong Polytechnic University\\
  Hong Kong SAR, China \\
  \And
  Qing Li \\
  Department of Computing\\
  The Hong Kong Polytechnic University\\
  Hong Kong SAR, China \\
}

\begin{document}

\maketitle

\begin{abstract}
Current research agents are moving beyond academic search toward literature synthesis and experiment planning. These workflows increasingly rely on literature retrieval systems, which serve as standalone systems or as components of research agents. However, existing scientific literature retrieval benchmarks mainly evaluate these methods using topical relevance or citation links as supervision, which cannot capture which prior methods are used for empirical comparison. To address this gap, we introduce \textbf{\textsc{Lineage}}, a benchmark for empirical lineage retrieval in AI scientific literature. An empirical lineage relation links a prior paper to a source paper that explicitly uses its method as an empirical comparison baseline. These relations form lineage graphs whose multi-hop neighborhoods provide retrieval ground truth for structural evaluation. To construct these graphs at scale, we develop \textbf{\textsc{Lineage}-Constructor}, which converts heterogeneous paper records and full-paper empirical baseline evidence into calibrated lineage graphs.  Built from 121,048 papers across 17 top AI venues from 2018 to 2025, \textsc{Lineage} contains 3,950 graph components, 67,502 paper nodes, 164,948 lineage edges, and 80,000 multi-facet retrieval queries. Experiments with 19 representative retrieval systems show that empirical lineage retrieval remains challenging. Performance drops sharply with lineage distance, while structural metrics reveal that isolated retrieval hits often fail to recover closely connected empirical comparison neighborhoods.
\end{abstract}

\input{1-introduction}
\input{2-related-work}
\input{3-construction}
\input{4-benchmark}
\input{5-results}
\input{6-conclusion}

\clearpage
\bibliographystyle{plain}
\bibliography{ref}

\clearpage
\appendix
\input{appendix/appendix}

\end{document}

%% file: 1-introduction.tex
\section{Introduction}
Large language models (LLMs) are increasingly being developed into research agents for automating scientific workflows~\cite{lu2026towards,lyu2026evoscientist}. These agents support different stages of research, including hypothesis formulation~\cite{sican,tangai}, literature synthesis~\cite{asai2026synthesizing,chen2026mars}, experiment automation~\cite{chenmlr,chenscienceagentbench,huang2024mlagentbench,toledoai}, and manuscript generation~\cite{song2026paperorchestra,wang2024autosurvey}. Across these stages, literature retrieval provides the evidence needed for downstream research. Retrieval methods can also serve as standalone systems for academic search. However, research-oriented retrieval often requires more than finding topically similar papers. For literature synthesis and experiment planning, researchers may need to identify which prior methods are used as empirical baselines and which later papers compare against a given method. These relations reveal a paper's empirical comparison context and support baseline discovery and contribution positioning. Recent studies further show that paper retrieval quality affects downstream reasoning and generation~\cite{pan2026paperscout,xu2026train}, motivating benchmarks that evaluate retrieval beyond topical relevance.

Existing scholarly retrieval benchmarks primarily build upon relevance-based or citation-oriented criteria. Relevance-based benchmarks use aspect-specific queries to retrieve papers matching a scientific statement~\cite{ajith2024litsearch,wang2023scientific}, but topical relevance does not imply empirical comparison. Two papers may study the same problem without being directly compared in experiments. Citation-oriented benchmarks instead use citation recommendation or prediction as supervision~\cite{he2025pasa,long2024recommending,press2024citeme,zheng2026should}, but citation relations are heterogeneous and often reflect background discussion, datasets, tools, or theoretical support. As a result, neither signal isolates which prior methods are explicitly used for empirical comparison. Existing benchmarks therefore cannot directly evaluate whether retrieval methods recover the empirical comparison context surrounding a paper.

\begin{figure}
    \centering
    \vspace{1em}
    \includegraphics[width=1\linewidth]{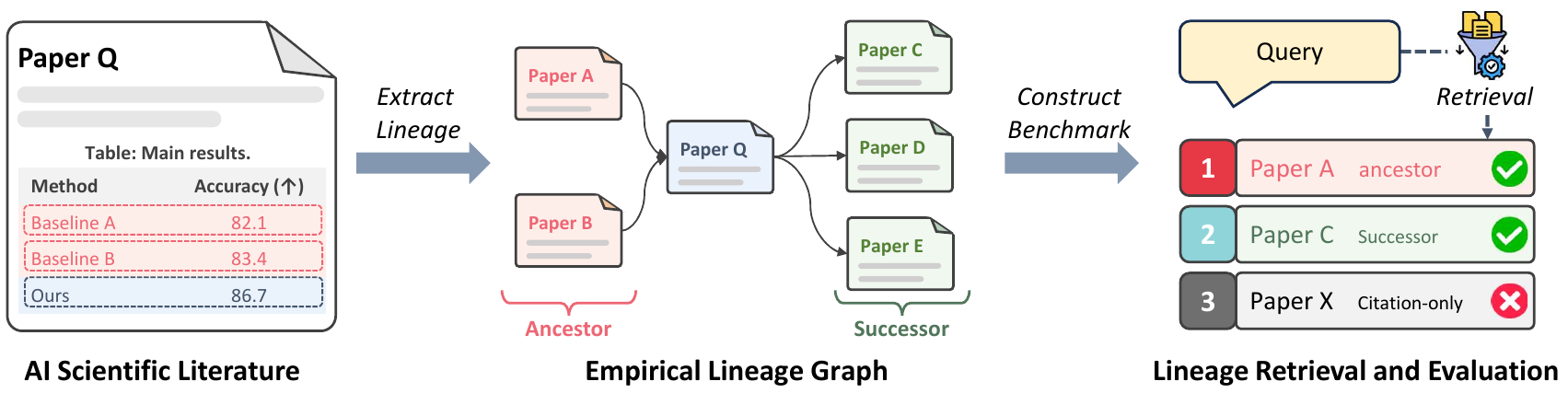}
    \caption{Illustration of empirical lineage relations. In \textsc{Lineage}, empirical baseline evidence induces directed lineage edges from baseline papers to source papers. Given a query paper, its ancestors and successors in the lineage graph define retrieval ground truth. These relations capture empirical comparison structure beyond generic topical or citation-based relevance.}
    \label{fig:lineage_illustration}
    \vspace{-2em}
\end{figure}

To address this gap, we propose \textbf{empirical lineage} as an auditable relation for literature retrieval. As illustrated in Figure~\ref{fig:lineage_illustration}, an empirical lineage relation links a prior paper to a source paper that explicitly uses its method as a baseline for empirical comparison. Each relation is grounded in explicit evidence. These relations capture the empirical reference frame used to evaluate and position a contribution. By linking these relations across papers, we formulate \textbf{empirical lineage retrieval} as the task to evaluate whether a retrieval system can recover papers in these comparison-based neighborhoods.

However, constructing empirical lineage graphs from scientific literature is non-trivial. 
\textbf{\ding{182}~Corpus Grounding:} Lineage retrieval requires an auditable retrieval space where each paper node and each lineage edge can be consistently identified. This is difficult because scientific papers are distributed across heterogeneous open-access sources with inconsistent coverage, metadata, and PDF availability. 
\textbf{\ding{183}~Evidence-to-Edge Ambiguity:} Unlike citation links, lineage relations must be identified from full-paper empirical evidence. Baseline mentions can appear in different sections and forms, making it difficult to distinguish true comparison targets from other citations.
\textbf{\ding{184}~Graph Suitability:} Extracted lineage relations do not directly form graphs suitable for retrieval evaluation. Some widely reused baselines can create overly broad connected regions even when individual comparison relations are limited. Such regions make multi-hop neighborhoods less suitable as retrieval ground truth.

To address these challenges, we propose \textbf{\textsc{Lineage}-Constructor}, an automatic pipeline for constructing empirical lineage graphs from scientific literature. The pipeline consists of three stages. 
\textbf{\ding{182}~Unified Literature Collector} builds a unified retrieval corpus from heterogeneous open-access sources, aligning metadata, validating PDFs, and normalizing paper records.
\textbf{\ding{183}~Evidence-Grounded Lineage Extraction} extracts lineage relations from empirical baseline evidence. Since such evidence can appear across different parts in varying formats, we design a dual-source extractor that provides complementary paper representations and enables cross-checking of outputs, improving coverage of dispersed baseline mentions while reducing spurious edges. We further annotate a human-annotated validation set, \textsc{Lineage}-HA, to guide extractor optimization and evaluate extraction quality. \textbf{\ding{184}~Graph Calibration} enforces temporal validity and regularizes abnormal graph connectivity, yielding calibrated lineage graphs for retrieval evaluation.

Based on these calibrated graphs, we build the \textbf{\textsc{Lineage}} benchmark for empirical lineage retrieval in artificial intelligence (AI) scientific literature. As summarized in Table~\ref{tab:benchmark_comparison}, \textsc{Lineage} is, to the best of our knowledge, the first large-scale literature retrieval benchmark using empirical lineage relations as retrieval ground truth. For each query paper, \textsc{Lineage} generates \textit{multi-facet queries} that describe the same paper from different retrieval intents, using its ancestor and successor papers in the lineage graph as retrieval ground truth. Beyond standard ranking metrics such as recall and nDCG, \textsc{Lineage} introduces structural metrics that measure whether retrieved papers recover coherent empirical comparison neighborhoods rather than isolated lineage hits.

The \textsc{Lineage} benchmark comprises \textbf{121,048 papers} from \textbf{17 top AI conference venues} published between 2018 and 2025. From this corpus, \textsc{Lineage}-Constructor constructs \textbf{3,950 directed lineage graph components}, encompassing 67,502 papers and 164,948 lineage edges. Based on these graphs, we generate \textbf{80,000 multi-facet retrieval queries}.
Experimental results highlight the substantial challenges in empirical lineage retrieval: 
(1) Even the strongest agentic retriever achieves only 31.87 Recall@20 and 9.92 Precision@20, indicating that current retrievers fail to comprehensively recover empirical lineage targets. 
(2) Performance degrades rapidly with lineage distance, with average Recall@20 decreasing from 28.42 for one-hop targets to 6.68 for three-hop targets. 
(3) Graph-based metrics further reveal that occasional retrieval hits do not imply recovery of lineage structure, as online search engines achieve non-trivial hit rates but much lower structural scores.

Our contributions are three-fold: 1) We propose \textbf{empirical lineage relations} as a ground-truth signal for benchmarking literature retrieval, enabling the evaluation of whether retrieval methods recover experiment-grounded empirical neighborhoods beyond topical relevance and citation connectivity. 2) We design \textbf{\textsc{Lineage}-Constructor}, an automatic pipeline that builds lineage graphs by extracting empirical baseline evidence from full papers. The pipeline integrates a unified literature collector, evidence-grounded lineage extraction, and temporal and structural lineage graph calibration to improve the quality of the constructed lineage graphs. 3) We introduce \textbf{\textsc{Lineage}}, a large-scale benchmark featuring multi-facet queries and structural metrics, to systematically evaluate representative retrieval systems, revealing substantial gaps in capturing empirical lineage and highlighting opportunities for future research.

\begin{table}[t]
\centering
\caption{Comparison with relevance-based and citation-oriented literature retrieval benchmarks.}
\label{tab:benchmark_comparison}
\resizebox{\linewidth}{!}{
\begin{tabular}{l|lccccc}
\toprule
\textbf{Benchmark} & \textbf{Ground Truth} &
\begin{tabular}[c]{@{}c@{}}\textbf{Multi-Facet}\\\textbf{Query}\end{tabular} &
\begin{tabular}[c]{@{}c@{}}\textbf{Automatic}\\\textbf{Construction}\end{tabular} &
\begin{tabular}[c]{@{}c@{}}\textbf{Lineage}\\\textbf{Relation}\end{tabular} &
\begin{tabular}[c]{@{}c@{}}\textbf{Multi-Hop}\\\textbf{Retrieval}\end{tabular} &
\begin{tabular}[c]{@{}c@{}}\textbf{Structural}\\\textbf{Metrics}\end{tabular} \\
\midrule
LitSearch & Relevant Paper & \xmark & \xmark & \xmark & \xmark & \xmark \\
CiteME & Citation & \cmark & \xmark & \xmark & \xmark & \xmark \\
DORIS-MAE & Relevant Paper & \cmark & \xmark & \xmark & \xmark & \xmark \\
CitationR & Citation & \xmark & \cmark & \xmark & \xmark & \xmark \\
PaSa & Citation & \xmark & \cmark & \xmark & \xmark & \xmark \\
CiteRAG & Citation & \cmark & \cmark & \xmark & \xmark & \xmark \\
\midrule
\textbf{\textsc{Lineage} (Ours)} & Lineage Relation & \cmark & \cmark & \cmark & \cmark & \cmark\\
\bottomrule
\end{tabular}}
\vspace{-1em}
\end{table}

%% file: 2-related-work.tex
\section{Related Work}
Academic research benchmarks evaluate AI systems across scholarly workflows, including document processing, scientific document understanding, literature retrieval, citation-oriented recommendation, scientific question answering, and broader research-agent evaluation~\cite{chenmlr,cohan2020specter,dasigi2021dataset,huang2024mlagentbench,lo2020s2orc,singh2023scirepeval,wadden2020fact,yuan2026magic}. 
Recent work further narrows this focus to literature retrieval, examining fine-grained scenarios such as aspect-based scientific queries, complex literature search, citation-grounded retrieval, and retrieval over textual or relational scientific knowledge bases~\cite{ajith2024litsearch,press2024citeme,shao2025scinetbench,wang2023scientific,wu2024stark,zheng2026should}.
Meanwhile, LLM-based research agents increasingly rely on literature retrieval to support research ideation, experiment planning, survey generation, manuscript writing, and academic search~\cite{he2025pasa,pan2026paperscout,sican,song2026paperorchestra,toledoai,wang2024autosurvey,wengcycleresearcher}. However, their evaluations typically emphasize topical relevance, citation-based association, evidence support, or downstream research-oriented generation. In contrast, \textsc{Lineage} targets \textbf{empirical lineage retrieval}. It uses explicit empirical baseline relations as retrieval ground truth.

%% file: 3-construction.tex
\section{\textsc{Lineage}-Constructor}
\begin{figure}[t]
    \centering
    \includegraphics[width=\linewidth]{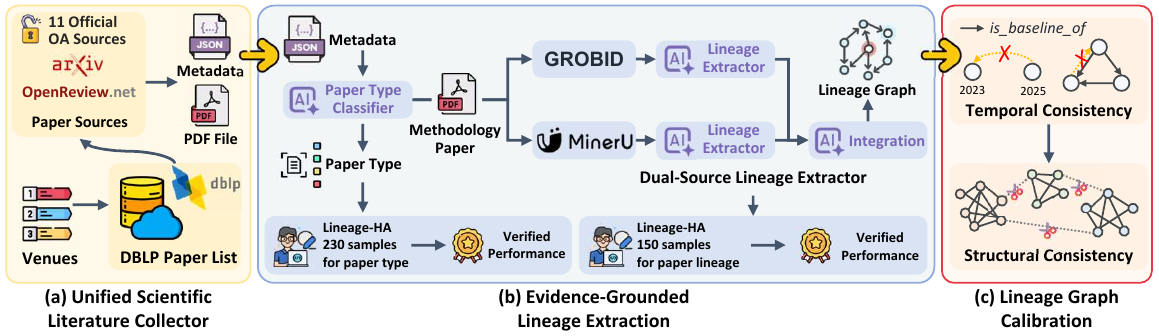}
    \caption{Overview of the \textsc{Lineage}-Constructor pipeline, consisting of (a) unified scientific literature collector, (b) evidence-grounded lineage extraction, and (c) lineage graph calibration.}
    \label{fig:pipeline}
    \vspace{-1em}
\end{figure}

The \textsc{Lineage}-Constructor pipeline involves three major modules: unified scientific literature collector, evidence-grounded lineage extraction, and lineage graph calibration, as illustrated in Figure~\ref{fig:pipeline}. These modules enable efficient automated construction with traceable quality, thereby significantly reducing construction costs, as discussed in \S\ref{app:cost}.

\subsection{Unified Scientific Literature Collector}
Constructing empirical lineage graphs requires literature records that are both comprehensive and consistently represented. Unlike citation-based retrieval benchmarks, which rely on a single paper, lineage retrieval needs broad coverage across AI venues, as each directed edge points from the baseline paper to the source paper that uses it as an empirical comparison baseline. Literature collection is, therefore, a core component of benchmark construction.

In practice, there is no single open-access source providing complete and uniform records. DBLP~\cite{dblp} offers reliable venue-year paper lists, but abstracts and PDFs often need to be fetched from other providers. These providers differ in metadata schema, naming conventions, URL formats, and file availability. Directly merging them may create inconsistent records and noisy classifiers, which can propagate errors into lineage extraction and graph construction.

We built a unified literature collector module to address this issue. The module integrates 11 complementary open-access sources. It first constructs a canonical venue-year paper list from DBLP and filters out records that are unlikely to be regular research papers, including tutorials, demos, workshops, and student abstracts. It then enriches each retained paper using venue-specific open-access sources. Official proceedings are prioritized when available. OpenReview~\cite{openreview} and arXiv~\cite{arxiv} are used to recover missing PDFs. The venue-specific source configuration is provided in \S\ref{app:pdf-acquistion}. We restrict the benchmark corpus to peer-reviewed conference papers, as they commonly provide explicit empirical comparisons and consistent full-paper evidence for lineage extraction.

After metadata enrichment, the module downloads and validates the corresponding PDFs before downstream parsing. A lightweight page-length filter removes short files that are unlikely to be regular papers. The module also performs diagnostic checks for suspicious metadata and PDFs, as detailed in \S\ref{app:metadata-quality}. These checks provide a practical repair interface and reduce the risk of noisy records propagating into the lineage construction process.

\subsection{Evidence-Grounded Lineage Extraction}
\label{sec:lineage-extraction}

Unlike citation links, which are explicitly marked and often available from structured metadata, empirical lineage relations must be identified from full-paper comparison evidence. These signals are distributed across papers and often use non-canonical method names, making direct rule-based construction unreliable. We therefore design an evidence-grounded lineage extraction module that progressively narrows the corpus, grounds extracted relations in paper content, and converts the retained relations into a corpus-grounded lineage graph.

\textbf{Paper Type Classifier.}
Regular papers in our corpora include various contribution types, such as methodology papers, benchmark papers, and surveys. Since benchmark and survey papers primarily build evaluation resources or synthesize existing work rather than proposing new technical methods, their citation relations are less suitable for constructing lineage graphs. We therefore first identify methodology papers for downstream lineage extraction. Specifically, we prompt \texttt{Qwen3.5-27B}\footnote{https://huggingface.co/Qwen/Qwen3.5-27B} with each paper's title and abstract. The prompt classifies papers by their primary contribution, classifying papers as \textit{methodology} only when the central contribution is a technical method rather than an evaluation resource or a synthesis of prior work. Detailed prompts are provided in \S\ref{app:graph-prompt}.

\textbf{Dual-Source Lineage Extractor.}
To robustly extract lineage relations from scientific papers, we propose a dual-source lineage extractor with candidate-restricted integration. Each PDF is parsed by two complementary tools: \texttt{GROBID}~\cite{grobid}, which produces semi-structured XML, and \texttt{MinerU}~\cite{wang2024mineru}, which produces markdown with paper-related structures. Rather than concatenating the two parser outputs, we prompt \texttt{Qwen3.5-27B} to \textbf{extract} candidate lineage relations from each source independently with the prompt presented in \S\ref{app:graph-prompt}. We then \textbf{integrate} the two candidate sets through a grounding step that re-checks each candidate against the paper content and retains only supported relations. This design uses parser-specific extraction to improve recall while limiting integration to evidence-based conflict resolution. The model also normalizes cited paper titles and removes duplicates.

\textbf{\textsc{Lineage}-HA.}
To evaluate the reliability of our automated extraction module, we construct \textsc{Lineage}-HA, a human-annotated validation set containing 230 papers annotated for paper type classification and 150 papers annotated for lineage relation extraction. The lineage-relation annotations contain 970 positive empirical lineage relations. It allows us to evaluate whether the module correctly filters methodology papers and extracts supported lineage relations from full papers. Details of \textsc{Lineage}-HA are provided in \S\ref{app:lineage-ha}, where the results show that our paper type classification and dual-source lineage extraction outperform their alternatives.

\textbf{Lineage Graph Construction.}
After extracting lineage relations from individual papers, we construct a corpus-grounded directed lineage graph $\mathcal{G}=(\mathcal{V},\mathcal{E})$ over the collected paper database. Each node $v \in \mathcal{V}$ denotes a paper in the collected corpus, and each directed edge $e=(v_b,v_s)\in\mathcal{E}$ indicates that the source paper $v_s$ explicitly uses the method from $v_b$ as an empirical comparison baseline. We retain only source--baseline pairs in which both papers can be matched to the collected corpus, so that every graph node belongs to the same retrieval space used for downstream evaluation. Nodes store basic paper metadata, while edges retain extraction provenance. The resulting graph encodes ``\verb|is_baseline_of|'' relations, with the reverse direction corresponding to ``\verb|uses_baseline|''.

\subsection{Lineage Graph Calibration}
To enhance the reliability of constructed lineage graphs, we employ temporal and structural consistency calibration to filter out abnormal relations. Temporal calibration enforces valid edge ordering. Structural calibration controls overly broad connectivity in multi-hop neighborhoods. Stage-wise calibration analysis is in \S~\ref{app:metadata-quality}.

\textbf{Temporal Consistency.}
We first calibrate the graph by enforcing temporal validity. A retained lineage edge must point from a baseline paper to a source paper published in the same year or later, since an empirical baseline cannot depend on future work. Edges that violate this ordering are removed as invalid lineage candidates. 

\textbf{Structural Consistency.}
We further calibrate the graph to make its connected regions suitable for lineage-based retrieval evaluation. Because widely reused baselines can create very large connected components, raw components can be too broad for defining informative multi-hop retrieval neighborhoods. We therefore apply a fixed component-level calibration procedure to split components classified as upper-tail outliers by the \texttt{IQR} rule~\cite{chatfield1986exploratory} into more coherent subgraphs using bridge pruning~\cite{tarjan1974note} and Louvain community partitioning~\cite{blondel2008fast}. Finally, we discard isolated nodes that no longer participate in any retained lineage relation. These operations produce lineage graphs with valid temporal ordering and controlled connectivity for retrieval evaluation.

%% file: 4-benchmark.tex
\section{Dataset}
Based on the calibrated lineage graphs constructed by \textsc{Lineage}-Constructor, we introduce \textsc{Lineage}, a benchmark for empirical lineage retrieval. It covers papers from 17 top AI venues from 2018 to 2025, including 121,048 papers with complete metadata records, as detailed in \S\ref{app:corpus-coverage}. After lineage extraction and calibration, we construct 80,000 multi-facet retrieval queries as the \textbf{\textsc{Lineage} benchmark}. It supports evaluation with both standard retrieval metrics and graph-structure metrics. 

Specifically, given a calibrated directed lineage graph $G=(V, E)$, \textsc{Lineage} evaluates each query paper $v_q\in V$ by generating multi-facet retrieval queries. A retrieval method $\mathcal{M}$ returns a ranked list $R_k(q,\mathcal{M})$ from our corpus $\mathcal{C}$ (\emph{i.e.,} the collected 121,048 papers), and the retrieved papers are compared with the empirical lineage ground-truth papers $Y_q$, which are derived from the ancestors and successors of $v_q$.

\begin{figure}[t]
\centering
\includegraphics[width=\linewidth]{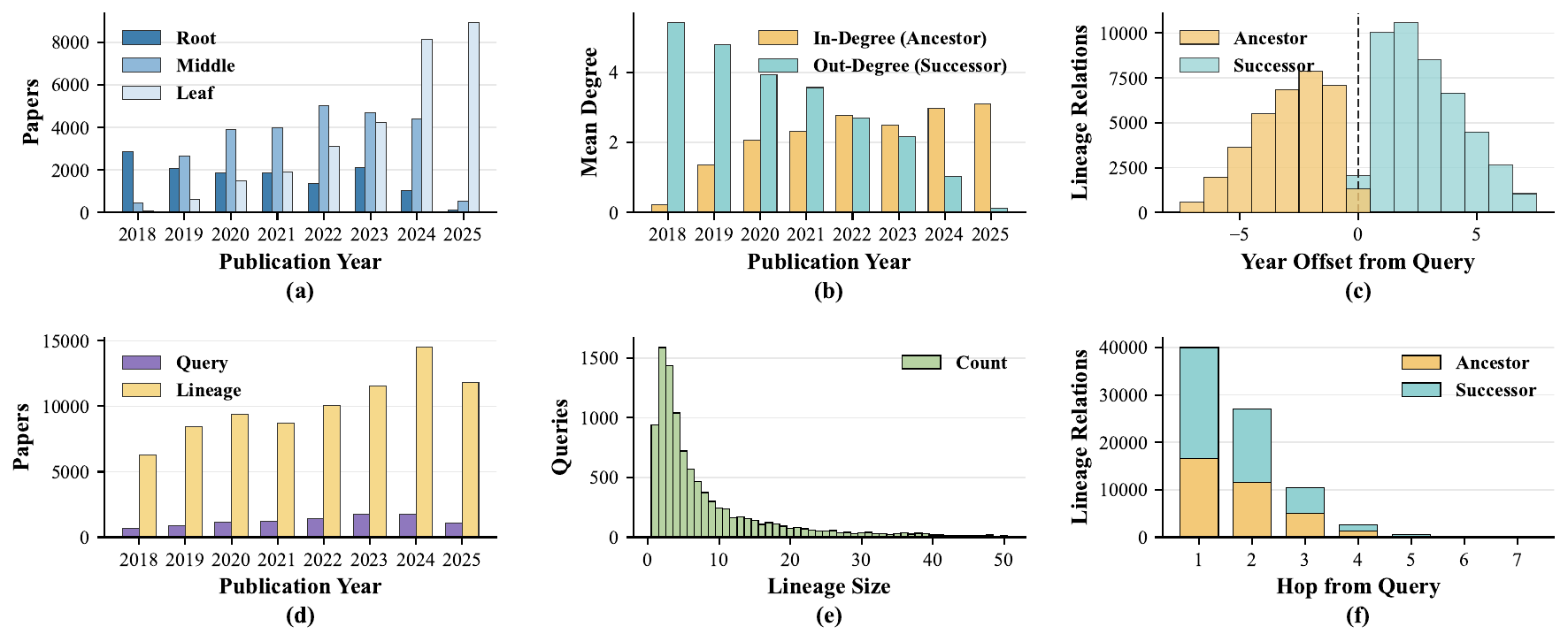}
\caption{Graph statistics (a-c) and query properties (d-f) in \textsc{Lineage}. 
(a) Publication-year distribution of root, middle, and leaf nodes.
(b) Average in-degree and out-degree by publication year.
(c) Year offsets between query papers and their ancestor or successor ground-truth papers.
(d) Yearly distribution of query papers and lineage papers.
(e) Distribution of ground-truth lineage set size per query.
(f) Distribution of ground-truth lineage papers by hop distance from the query paper.
}
\label{fig:lineage_graph_stat}
\vspace{-1.0em}
\end{figure}

\subsection{Lineage Graph}
\begin{wraptable}{r}{0.38\textwidth}
\vspace{-1.0em}
\centering
\caption{\textsc{Lineage} and query statistics.}
\label{tab:lineage_statistics}
\footnotesize
\resizebox{\linewidth}{!}{\begin{tabular}{c|l|r}
\toprule
\textbf{Domain} & \textbf{Statistic} & \textbf{Value} \\
\midrule
\multirow[c]{6}{*}{\makecell[c]{Lineage\\Graph}}
& Component & 3,950 \\
& \#Node (\textit{Paper}) & 67,502 \\
& ~~~~~\#Root & 13,315 \\
& ~~~~~\#Leaf & 28,488 \\
& ~~~~~\#Middle & 25,699 \\
& \#Edge (\textit{Lineage}) & 164,948 \\
\midrule
\multirow[c]{5}{*}{\makecell[c]{Multi-\\Facet\\Queries}}
& \#Query papers  & 10,000 \\
& \#Query         & 80,000 \\
& ~~~~~\#Root     & 26,672 \\
& ~~~~~\#Leaf     & 26,664 \\
& ~~~~~\#Middle   & 26,664 \\
& Lineage/Query   & 8.1 \\
\bottomrule
\end{tabular}}
\vspace{-2.0em}
\end{wraptable}

In \textsc{Lineage}, each calibrated lineage graph contains papers connected through empirical baseline relations. Nodes correspond to papers, and directed edges indicate explicit empirical comparison relations. We orient each edge from a baseline paper to the source paper that uses its method for empirical comparison. Under this direction, ancestors trace earlier papers connected through one or more baseline relations, while successors trace later papers connected in the opposite direction. As shown in Table~\ref{tab:lineage_statistics}, \textsc{Lineage} contains 3,950 lineage graphs, 67,502 paper nodes, and 164,948 lineage edges. The graph collection includes 13,315 root nodes and 28,488 leaf nodes, covering both starting-point and terminal papers under the current observation window.

Figures~\ref{fig:lineage_graph_stat}(a)--(c) illustrate the temporal structure of the constructed lineage graph. At the graph level, root nodes concentrate in earlier years, whereas leaf nodes become more common in recent years, as shown in Figure~\ref{fig:lineage_graph_stat}(a). This node-level tendency is complemented by the degree statistics in Figure~\ref{fig:lineage_graph_stat}(b), where earlier papers have larger successor out-degree and recent papers have larger ancestor in-degree.  The relation timing gives a more direct view of this ordering. Ancestors lie mainly before the paper under consideration, and successors mainly after it, as reported in Figure~\ref{fig:lineage_graph_stat}(c). These distributions show that the constructed empirical lineage graphs follow the expected temporal ordering. Additional statistics and cases are provided in \S\ref{app:lineage-structure} and \S\ref{app:graph-case}.

\subsection{Multi-Facet Queries}
\label{sec:multi-facet-queries}
Common literature search behaviors are heterogeneous across users and retrieval methods. To cover these scenarios, we generate eight multi-facet retrieval queries for each query paper by prompting an LLM to extract and reformulate salient information from the title and abstract by the prompt illustrated in \S\ref{app:query-prompt}. Specifically, the \bi{title} query uses the paper title directly and targets exact paper lookup, while the \bi{title-context} query concatenates the title and abstract to provide more detailed information. The \bi{topical keyword} query extracts a short set of keywords, while the \bi{topical semantic} query rephrases the problem as a concise query. We further generate keywords and semantic variants for two additional facets. The contextual facet focuses on the background, yielding \bi{context keyword} and \bi{context semantic} queries. The methodological facet focuses on the technical mechanism, yielding \bi{method keyword} and \bi{method semantic} queries. \S\ref{app:query-case} illustrates the cases for each kind of query.

To reduce noisy queries, we use \texttt{Jina-Embedding-v5-0.6B}~\cite{akram2026jina} to retrieve the query paper from which each query is generated. If the query paper is not ranked among the top 20 results, the query is regenerated. As shown in Table~\ref{tab:lineage_statistics}, the selected query papers are balanced across root, leaf, and middle nodes. 
The query set maintains broad temporal coverage, with its yearly distribution closely following that of lineage papers in Figure~\ref{fig:lineage_graph_stat}(d).  The retrieval ground-truth sets are also heterogeneous, as their size follows a long-tailed distribution in Figure~\ref{fig:lineage_graph_stat}(e). Beyond variable target sizes, the hop distribution spans multiple graph distances, as shown in Figure~\ref{fig:lineage_graph_stat}(f).
It covers both direct empirical comparison neighbors and papers connected through longer lineage paths, enabling evaluation across a broader structural range. We further report the statistics of queries and ground-truth in \S\ref{app:query-distribution}.

\subsection{Evaluation Metrics}
In \textsc{Lineage}, we evaluate retrieval using both standard and graph-structural metrics. 

\textbf{Retrieval.} Standard retrieval metrics, including \underline{\textbf{R}}ecall@$k$, \underline{\textbf{P}}recision@$k$, \underline{\textbf{H}}it Rate@$k$, \underline{\textbf{N}}DCG@$k$, and \underline{\textbf{M}}RR@$k$, measure whether lineage ground-truth papers are retrieved and whether they appear in high-ranking positions. 
Since the ground truth is an unordered set of lineage papers, \underline{\textbf{N}}DCG@$k$ is computed with binary relevance. 

\textbf{Structural.} Since standard metrics do not capture retrieval performance over the lineage graph structure, we introduce several structural metrics for lineage retrieval. Graph distance reflects the length of the empirical comparison chain connecting a target paper to the query paper. \underline{\textbf{H}}op-\underline{\textbf{W}}eighted \underline{\textbf{R}}ecall@$k$ assigns larger weights to papers connected through shorter paths. \underline{\textbf{D}}irect \underline{\textbf{R}}ecall@$k$ focuses on one-hop lineage neighbors, which correspond to explicit empirical baseline relations. \underline{\textbf{G}}raph-weighted \underline{\textbf{n}}DCG@$k$ further evaluates whether structurally closer lineage papers are ranked higher.

%% file: 5-results.tex
\begin{table}[t]
\centering
\caption{Evaluation results of representative retrieval methods on \textsc{Lineage}. The best result in each metric is highlighted in \textbf{bold}, and the second-best result is \underline{underlined}.}
\label{tab:retrieval_results}
\resizebox{0.94\linewidth}{!}{
\begin{tabular}{cl|rrrrrrrr}
\toprule
\multirow{2}{*}{\textbf{Category}} 
& \multirow{2}{*}{\textbf{Retrieval Method}} 
& \multicolumn{5}{c}{\textbf{Retrieval Metric}} 
& \multicolumn{3}{c}{\textbf{Structural Metric}} \\
\cmidrule(lr){3-7}
\cmidrule(lr){8-10}
& & \textit{R}@20 & \textit{P}@20 & \textit{H}@20 & \textit{N}@20 & \textit{M}@20 & \textit{HWR}@20 & \textit{DR}@20 & \textit{GN}@20 \\
\midrule
\rowcolor{gray!20} \multicolumn{2}{l|}{\textit{w/} \textsc{Lineage} \textit{Corpus} (Avg.)} 
& 24.82 & 7.34 & 54.95 & 19.54 & 27.34 & 26.01 & 28.42 & 19.70 \\
\midrule
\multirow{2}{*}{\makecell[c]{Sparse\\Retrieval}}
& BM25
& 22.06 & 6.48 & 52.81 & 17.56 & 25.90 & 23.28 & 25.74 & 17.88 \\
& TF-IDF
& 20.42 & 6.02 & 49.90 & 15.81 & 22.95 & 21.48 & 23.64 & 16.00 \\
\midrule

\multirow{8}{*}{\makecell[c]{Dense\\Retrieval}}
& Contriever
& 18.20 & 5.24 & 46.91 & 14.29 & 21.52 & 19.18 & 21.19 & 14.48 \\
& BGE-M3 (0.6B)
& 10.71 & 3.03 & 32.54 & 8.30 & 13.43 & 11.24 & 12.34 & 8.37 \\
& Jina-Embedding-v5 (0.6B)
& 26.26 & 7.58 & 57.14 & 20.50 & 28.48 & 27.64 & 30.46 & 20.83 \\
& Qwen3-Embedding (0.6B)
& 25.96 & 7.69 & 56.74 & 20.40 & 28.31 & 27.34 & 30.18 & 20.71 \\
& Qwen3-Embedding (4B)
& \underline{30.64} & \underline{8.92} & \underline{61.88} & \underline{23.90} & \underline{31.92} & \underline{32.10} & \underline{35.09} & \underline{24.10} \\
& Qwen3-Embedding (8B)
& 29.35 & 8.52 & 60.32 & 22.78 & 30.54 & 30.60 & 33.18 & 22.80 \\
& NV-Embedding-v2 (8B)
& 29.76 & 8.87 & 61.21 & 23.14 & 30.92 & 31.28 & 34.38 & 23.41 \\
& ReasonIR (8B)
& 25.38 & 7.35 & 57.03 & 19.72 & 27.71 & 26.60 & 29.11 & 19.88 \\
\midrule
\multirow{4}{*}{\makecell[c]{Agentic\\Retrieval}}
& Pasa
& 20.41 & 6.07 & 49.50 & 17.06 & 25.74 & 21.41 & 23.47 & 17.27 \\
& AgentIR$^\diamond$
& 27.61 & 8.29 & 60.12 & 22.15 & 31.31 & 28.85 & 31.43 & 22.28 \\
& AgentIR-Qwen3-Embedding$^\diamond$
& 28.78 & 8.82 & 59.25 & 22.93 & 30.98 & 29.88 & 31.81 & 22.82 \\
& Tongyi-DeepResearch$^\diamond$
& \textbf{31.87} & \textbf{9.92} & \textbf{63.88} & \textbf{24.99} & \textbf{33.09} & \textbf{33.23} & \textbf{35.89} & \textbf{24.99} \\

\midrule
\rowcolor{gray!20} \multicolumn{2}{l|}{\textit{w/} \textit{Online Sources} (Avg.)}
& 6.69 & 1.90 & 24.95 & 5.62 & 10.31 & 6.89 & 5.59 & 4.33 \\
\midrule
\multirow{2}{*}{\makecell[c]{Search\\Engine}}
& Google Search$^\diamond$
& 9.64 & 0.40 & 32.37 & 6.64 & 10.62 & 9.99 & 2.23 & 1.42 \\
& Google Scholar$^\diamond$
& 7.57 & 2.00 & 23.62 & 5.23 & 8.11 & 8.01 & 8.78 & 5.14 \\
\midrule

\multirow{3}{*}{\makecell[c]{LLM\\Search}}
& GPT-5.4$^\diamond$ 
& 7.24 & 3.30 & 29.13 & 7.12 & 13.56 & 7.42 & 7.87 & 6.65 \\
& Gemini-3.1$^\diamond$ 
& 4.47 & 1.97 & 21.50 & 4.58 & 10.35 & 4.51 & 4.46 & 4.23 \\
& DeepSeek-V3.2$^\diamond$ 
& 4.52 & 1.82 & 18.12 & 4.55 & 8.92 & 4.51 & 4.59 & 4.23 \\
\bottomrule
\end{tabular}}
\vspace{2pt}
\noindent
\parbox{0.95\textwidth}{
\hangindent=1em
\hangafter=1
\footnotesize $^\diamond$ Due to the rate limit of the API, we evaluate these methods on a sampled subset. The details of the subset and other methods' performance on it are presented in \S\ref{app:lineage-800}.
}
\vspace{-1em}
\end{table}

\section{Results}
In experiments, we evaluate 19 retrieval methods on \textsc{Lineage} under two settings. The corpus-based setting includes 14 methods that retrieve from the fixed \textsc{Lineage} corpus, covering two sparse retrievers based on lexical matching, eight dense retrievers based on embedding similarity, and four agentic retrieval methods based on iterative search and candidate selection. The online-source setting includes five methods, covering two search engines and three LLMs enforced by online search. Detailed model choices and implementation settings are provided in \S\ref{app:retriever_settings}.

The corpus-based baselines span three retrieval paradigms. Sparse retrievers include BM25~\cite{robertson2009probabilistic} and TF-IDF~\cite{sparck1972statistical}, which rely on lexical matching. Dense retrievers include Contriever~\cite{izacard2022unsupervised}, BGE-M3~\cite{chen2024m3}, Jina-Embedding-v5~\cite{akram2026jina}, Qwen3-Embedding (0.6B, 4B, and 8B)~\cite{zhang2025qwen3}, NV-Embedding-v2~\cite{lee2025nvembed}, and ReasonIR~\cite{shao2025reasonir}, which rank papers by embedding similarity. Agentic retrievers include PaSa~\cite{he2025pasa}, AgentIR and AgentIR-Qwen3-Embedding~\cite{chen2026agentir}, and Tongyi-DeepResearch~\cite{team2025tongyi}, which iteratively reformulate queries and select candidates.

\subsection{Main Results}
Table~\ref{tab:retrieval_results} reveals several findings about the difficulty of retrieving lineage-related papers and recovering empirical lineage structure in AI scientific literature.

\textbf{Lineage retrieval remains challenging.}
The main results show that empirical lineage retrieval remains far from solved.
Even the strongest method, Tongyi-DeepResearch, achieves only 31.87 R@20 and 9.92 P@20. Across categories, sparse retrievers provide competitive lexical baselines, dense retrievers improve performance with stronger semantic representations, and agentic retrievers achieve the best overall results. 
However, none of these categories retrieves lineage papers comprehensively. 
It indicates that current retrieval methods can sometimes identify relevant lineage papers, but still struggle to capture the full set of papers connected through empirical comparison associated with a query paper.

\textbf{Agentic retrieval improves performance but remains insufficient.}
Although agentic retrieval achieves the best overall performance, Pasa performs worse than BM25 on the full benchmark. These results suggest that current academic search agents primarily improve query reformulation and candidate selection, but still lack explicit mechanisms for tracing empirical comparison relations.

\textbf{Structural metrics reveal failures hidden by standard retrieval metrics.}
The structural metrics expose retrieval failures that are not fully captured by standard ranking metrics. For example, Google Search obtains a non-trivial H@20, but its DR@20 and GN@20 remain very low. This shows that occasional hits do not imply recovery of the local empirical comparison structure.

\begin{wrapfigure}{r}{0.42\textwidth}
\centering
\vspace{-1.0em}
\includegraphics[width=1\linewidth]{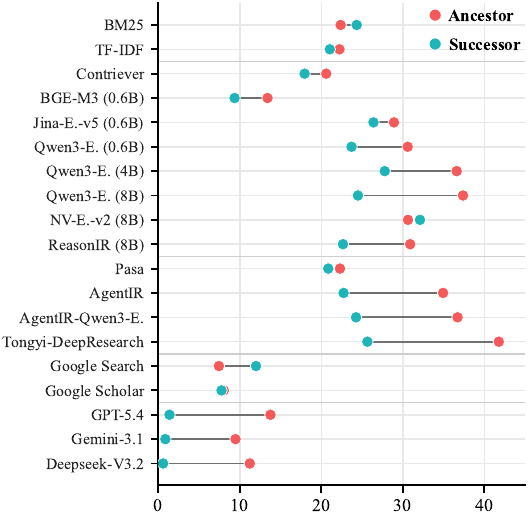}
\caption{Recall@20 for ancestor and successor retrieval across methods.}
\label{fig:ancestor-successor-recall}
\vspace{-1.5em}
\end{wrapfigure}
\textbf{A well-crafted paper source is necessary for lineage retrieval.}
The gap between corpus-based and online-source retrieval is substantial, with average R@20 dropping from 24.82 to 6.69 and average DR@20 dropping from 28.42 to 5.59. Although online-source methods are restricted to academic sources and the same publication years, they still struggle to consistently retrieve lineage papers. This suggests that lineage retrieval relies on a well-crafted paper source, highlighting the need for a clean, comprehensive, and stable data interface for evaluating empirical lineage retrieval.

\textbf{Ancestor retrieval is easier than successor retrieval.}
Figure~\ref{fig:ancestor-successor-recall} shows that most methods achieve higher R@20 on ancestors than successors. Dense retrievers exhibit noticeable variation in directional preference. The gap is especially large for agentic and LLM-based methods, suggesting that they struggle to identify later papers that use a given method as an empirical baseline.

\subsection{Retrieval Performance across Query Types}
\label{sec:query-type-exp}
\begin{figure}[t]
\centering
\begin{minipage}{0.55\linewidth}
    \centering
    \includegraphics[width=\linewidth]{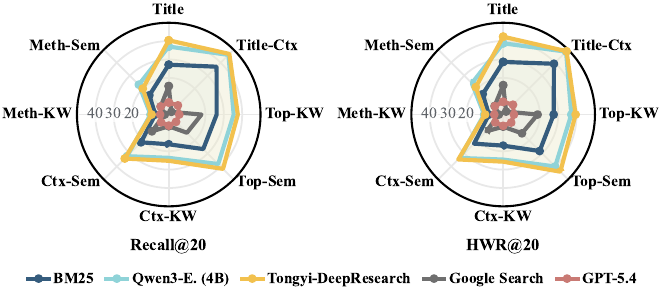}
    \caption{Methods' performance over query types.}
    \label{fig:query-recall}
\end{minipage}
\hfill
\begin{minipage}{0.42\linewidth}
    \centering
    \resizebox{0.86\linewidth}{!}{
    \includegraphics[width=\linewidth]{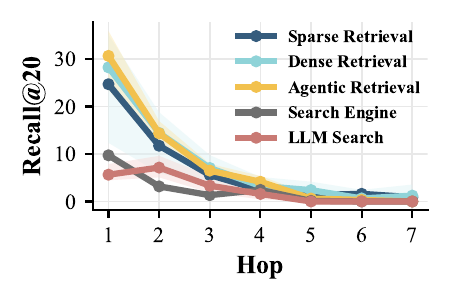}}
    \caption{Recall@20 across lineage hops.}
    \label{fig:hop-recall}
\end{minipage}
\vspace{-1em}
\end{figure}

To investigate how query representations affect lineage retrieval, we report category-level \textit{R}@20 and \textit{HWR}@20 across the eight query types in Table~\ref{tab:retrieval_query_results}. Method-level results are provided in \S\ref{app:query_type_results}. Retrieval performance varies substantially across query types. \textit{Title-Ctx} performs best for sparse, dense, and agentic retrievers, and also achieves the best result for LLM search. This suggests that combining the paper title with abstract-level context provides both paper-specific anchors and sufficient semantic information for retrieving lineage-related papers. Topical queries are also effective. \textit{Top-Sem} ranks second for dense and agentic retrievers, while \textit{Top-KW} performs best for search engines. In contrast, context-only and method-only queries are generally weaker, with \textit{Meth-KW} performing worst across categories. Furthermore, Figure~\ref{fig:query-recall} shows distinct query sensitivity across retrieval paradigms. BM25, Qwen3-E. (4B), and Tongyi-DeepResearch perform best on \textit{Title-Ctx} and \textit{Top-Sem}, while Google Search favors \textit{Top-KW}, suggesting that web search benefits more from concise topical keywords than from longer paper-level descriptions. GPT-5.4 remains weak and relatively insensitive to query formulation, indicating that query reformulation alone does not substantially improve its lineage retrieval capability. These results show that query formulation remains an important bottleneck even for stronger retrieval methods.

\begin{table}[t]
\centering
\caption{Performance of retrieval categories across different query types. For each category, the best query type is highlighted in \textbf{bold}, and the second-best query type is \underline{underlined}.}
\label{tab:retrieval_query_results}
\resizebox{0.95\linewidth}{!}{\begin{tabular}{cl|rrrrrrrr}
\toprule
\textbf{Category} & \textbf{Metric} & \textbf{\textit{Title}} & \textbf{\textit{Title-Ctx}} & \textbf{\textit{Top-KW}} & \textbf{\textit{Top-Sem}} & \textbf{\textit{Ctx-KW}} & \textbf{\textit{Ctx-Sem}} & \textbf{\textit{Meth-KW}} & \textbf{\textit{Meth-Sem}} \\
\midrule
\multirow{2}{*}{\makecell[c]{Sparse\\Retrieval}} & \textit{R}@20 
& \underline{25.60} & \textbf{36.58} & 24.66 & 25.41 & 15.52 & 20.75 & 7.03 & 14.37 \\
&  \textit{HWR}@20 
& \underline{26.96} & \textbf{38.79} & 25.95 & 26.82 & 16.19 & 21.71 & 7.42 & 15.18 \\
\midrule
\multirow{2}{*}{\makecell[c]{Dense\\Retrieval}} & \textit{R}@20 
& 29.44 & \textbf{39.71} & 28.19 & \underline{30.25} & 18.18 & 25.25 & 7.42 & 17.80 \\
&  \textit{HWR}@20 
& 30.96 & \textbf{41.97} & 29.61 & \underline{31.84} & 18.88 & 26.28 & 7.78 & 18.69 \\
\midrule
\multirow{2}{*}{\makecell[c]{Agentic\\Retrieval}} & \textit{R}@20 
& 34.57 & \textbf{38.92} & 33.46 & \underline{35.53} & 22.00 & 30.39 & 7.73 & 17.19 \\
&  \textit{HWR}@20 
& 36.20 & \textbf{40.84} & 35.00 & \underline{37.27} & 22.65 & 31.31 & 8.04 & 18.05 \\
\midrule
\multirow{2}{*}{\makecell[c]{Search\\Engine}} & \textit{R}@20 
& 8.27 & 1.02 & \textbf{16.31} & \underline{13.41} & 8.45 & 9.97 & 4.00 & 7.38 \\
&  \textit{HWR}@20 
& 8.51 & 1.14 & \textbf{17.36} & \underline{14.16} & 8.64 & 10.04 & 4.22 & 7.92 \\
\midrule
\multirow{2}{*}{\makecell[c]{LLM\\Search}} & \textit{R}@20 
& \underline{6.73} & \textbf{7.46} & 6.15 & 4.98 & 5.51 & 5.05 & 3.05 & 4.33 \\
&  \textit{HWR}@20 
& \underline{6.90} & \textbf{7.62} & 6.13 & 5.16 & 5.61 & 5.02 & 3.01 & 4.41 \\
\bottomrule
\end{tabular}}

\end{table}

\subsection{Retrieval Performance across Lineage Hops}
To examine whether retrieval methods can trace lineage papers beyond direct empirical comparison relations, we further group ground-truth papers by their shortest-path distance from the query paper and report category-level R@20 in Figure~\ref{fig:hop-recall}. Method-level results are provided in \S\ref{app:hop_results}. It shows that lineage distance substantially increases retrieval difficulty. The retrieval performance decreases rapidly as hop distance increases. These results highlight a key limitation of current agentic retrieval methods. While query reformulation and candidate selection improve their local retrieval performance, they do not explicitly traverse empirical lineage across papers. These results suggest that future agentic retrieval methods may benefit from explicit lineage expansion or lineage-aware supervision to better trace long-range empirical comparison chains.

%% file: 6-conclusion.tex
\section{Conclusion}
In this paper, we introduced \textsc{Lineage}, a benchmark for evaluating empirical lineage retrieval in AI scientific literature. 
\textsc{Lineage} uses empirical lineage relations as retrieval ground truth and evaluates whether retrieval methods can identify empirically connected papers. Our experiments show that existing retrieval methods still struggle to comprehensively retrieve lineage papers, highlighting the challenge of empirical lineage retrieval for future literature retrieval systems.

%% file: appendix/appendix.tex
\input{appendix/limitations}
\input{appendix/ex-related-work}
\input{appendix/evaluation-settings}
\input{appendix/additional-results}
\input{appendix/lineage-ha}
\input{appendix/paper-collection}
\input{appendix/statistics}
\input{appendix/case}
\input{appendix/prompts}

%% file: appendix/limitations.tex
\section{Limitations and Future Work}
\label{app:limitations}
\textsc{Lineage} is designed as a benchmark for empirical lineage retrieval grounded in empirical comparison baselines. The current release focuses on AI conference papers from 2018 to 2025, which provides a controlled and auditable setting for evaluating empirical lineage retrieval. A natural extension is to include journal papers and publications from other domains. However, such expansion depends on the licensing conditions of the underlying publications and on whether other domains provide mature and openly accessible APIs that support large-scale and reproducible data collection.

The benchmark focuses on paper-level retrieval over lineage graphs. 
Future work can build on \textsc{Lineage} to support more advanced research-agent settings, including interactive lineage exploration, lineage-based question answering, survey generation, and contribution positioning. These extensions would allow researchers to evaluate not only whether a system retrieves empirical lineage papers, but also whether it can use the recovered empirical comparison structure to support scientific reasoning and discovery.

\paragraph{Broader Impact.}
\textsc{Lineage} is intended to support more reliable literature retrieval and research-agent evaluation by distinguishing empirical lineage from topical or citation-based relevance. Its positive impact lies in helping researchers better compare retrieval methods and build research agents. A potential negative impact is that automatically constructed lineage graphs may be over-interpreted as complete or authoritative maps of scientific influence. This risk is mitigated by defining lineage relations narrowly as empirical baseline relations, reporting extraction and calibration procedures, and releasing the benchmark as an evaluation resource rather than a deployed decision system.

%% file: appendix/ex-related-work.tex
\section{Extended Related Work}
\label{app:ex-related-work}
\textbf{Academic research benchmarks} evaluate whether AI systems can support specific stages of academic research or complete research workflows. Early benchmarks and corpora provided evaluation resources for scholarly document processing~\cite{lo2020s2orc}, citation-aware representation and recommendation~\cite{cohan2020specter}, and broader scientific document understanding~\cite{singh2023scirepeval}. Recent benchmarks move toward more realistic scientific retrieval and evidence-grounded evaluation. DORIS-MAE evaluates aspect-based scientific queries~\cite{wang2023scientific}, LitSearch studies complex literature-retrieval questions~\cite{ajith2024litsearch}, and STARK and SciNetBench evaluate retrieval over textual and relational scientific knowledge bases~\cite{shao2025scinetbench,wu2024stark}. Citation-oriented benchmarks evaluate local citation recommendation, reviewer-identified missing citations, citation attribution, and RAG-based citation prediction~\cite{kang2018dataset,long2024recommending,press2024citeme,yuan2026mkg,zheng2026should}, while scientific QA benchmarks measure claim verification, full-paper question answering, and multi-document or multimodal evidence aggregation~\cite{dasigi2021dataset,lee2023qasa,li2024m3sciqa,pramanick2024spiqa,wadden2020fact,jiang2025qa}. Research-agent benchmarks further evaluate broader academic workflows, including report generation, experiment execution, and data analytics~\cite{chenmlr,chenbrowsecomp,chenscienceagentbench,huang2024mlagentbench,jiang2025hibench,majumder2024discoverybench}. Literature retrieval benchmarks evaluate whether agents can retrieve, rank, and synthesize relevant publications for downstream research tasks~\cite{asai2026synthesizing,he2025pasa,pan2026paperscout}. 
However, existing paper retrieval benchmarks mainly evaluate topical relevance, citation grounding, or evidence support, while overlooking empirical lineage relations among papers.

\textbf{LLM-based research agents} have recently received growing attention for automating the full research lifecycle, including idea generation, data collection, experiment execution, method optimization, iterative trial planning, and research evaluation~\cite{chen2026mars, lu2026towards, lyu2026evoscientist, wu2026datamart}. Recent work on research ideation showed the promise of automated idea generation in terms of feasibility and diversity~\cite{sican}. Studies on experiment execution further formulated AI research as a search process over runnable solutions, with an emphasis on search strategies, operators, and feedback-based exploration~\cite{toledoai, wengcycleresearcher, wu2026llm}. Beyond ideation and experimentation, research agents have also been developed to support research writing, survey generation, and figure generation~\cite{song2026paperorchestra,wang2024autosurvey,yang2026multimodal,zhu2026autofigure}. Literature retrieval agents further supported research workflows by retrieving, ranking, and synthesizing relevant publications for downstream research tasks~\cite{ asai2026synthesizing, he2025pasa, pan2026paperscout}. Recent studies investigated how multi-round retrieval agents can be improved through prompt design, reward modeling, and policy optimization~\cite{xu2026train}, highlighting the importance of literature retrieval in research agents.

%% file: appendix/evaluation-settings.tex
\section{Evaluation Settings}
\label{app:eval_settings}
Our source code and benchmark dataset are available at the anonymous GitHub repository: \url{https://anonymous.4open.science/r/Lineage-Benchmark-41F8}. Unless otherwise specified, we use a default random seed of 42. For both local and online LLMs, we use deterministic decoding by setting \texttt{temperature=0} and \texttt{do\_sample=False} to improve reproducibility.
Furthermore, the intermediate outputs produced by search engines and LLM-based search agents have been stored and will be released for reproduction.

\subsection{Metrics}
\label{app:structural_metrics}
The retrieval metrics evaluate the overall ability of a retriever to recover the unordered lineage ground-truth papers for each query paper and place them near the top of the ranking, without considering their graph distance or relation type.
\textbf{Recall}@$k$ quantifies how much of the lineage ground-truth papers are covered by the top-$k$ retrieved papers.
\textbf{Precision}@$k$ captures the fraction of top-$k$ retrieved papers that belong to the lineage ground-truth papers.
\textbf{Hit Rate}@$k$ indicates whether at least one lineage ground-truth paper appears in the top-$k$ retrieved results.
\textbf{nDCG}@$k$ characterizes how well lineage ground-truth papers are ranked near the top, using binary relevance because the ground truth is unordered.
\textbf{MRR}@$k$ reflects how early the first lineage ground-truth paper appears within the top-$k$ ranking.

We further introduce structural metrics to evaluate whether retrieved lineage papers are structurally close to the query paper in the lineage graph. Given a directed lineage graph $G=(V, E)$, each node $v\in V$ denotes a paper, and each directed edge $e=(v_b,v_s)\in E$ indicates that the baseline paper $v_b$ is a predecessor of the source paper $v_s$. 
For a query paper $v_q\in V$, let $Y_q\subseteq V$ denote its lineage ground-truth papers, and let $\mathbf{R}_k(v_q)=(r_1,\ldots,r_k)$ denote the ranked top-$k$ retrieval list. 
We use $R_k(v_q)=\{r_1,\ldots,r_k\}$ to denote the corresponding retrieved set. 
Since $Y_q$ may contain both predecessors and successors of $v_q$, we compute graph distance on the undirected version of $G$, denoted by $G^u$. 
Let $\mathrm{hop}(v_q,v)$ denote the shortest-path distance between $v_q$ and $v$ in $G^u$.

\paragraph{Hop-Weighted Recall@$k$.}
Hop-Weighted Recall@$k$ measures whether the retrieved results cover lineage papers that are structurally close to the query paper. 
It assigns larger weights to papers with smaller graph distances. 
Formally, Hop-Weighted Recall@$k$ is defined as,
\[
\mathrm{HWR@}k(v_q) =
\frac{\sum_{v\in R_k(v_q)\cap Y_q} \frac{1}{\mathrm{hop}(v_q,v)}}
{\sum_{v\in Y_q} \frac{1}{\mathrm{hop}(v_q,v)}}.
\]

\paragraph{Direct Recall@$k$.}
Direct Recall@$k$ focuses on the most immediate lineage relations. 
It measures the fraction of one-hop ground-truth neighbors that appear in the top-$k$ retrieved results. 
The one-hop ground-truth set is defined as,
\[
Y_q^{(1)} = \{v\in Y_q\mid \mathrm{hop}(v_q,v)=1\}.
\]
Accordingly, Direct Recall@$k$ is computed as,
\[
\mathrm{DR@}k(v_q) =
\frac{|R_k(v_q)\cap Y_q^{(1)}|}{|Y_q^{(1)}|}.
\]

\paragraph{Graph-weighted nDCG@$k$.}
Graph-weighted nDCG@$k$ evaluates ranking quality under a graph structure. 
It uses hop-based graded relevance, so that closer lineage papers receive larger gains when ranked higher. 
For each retrieved paper $r_i$, its graph-weighted gain is defined as,
\[
\mathrm{gain}(r_i) =
\begin{cases}
\frac{1}{\mathrm{hop}(v_q,r_i)}, & r_i\in Y_q,\\
0, & r_i\notin Y_q.
\end{cases}
\]
The resulting graph-weighted DCG is computed as,
\[
\mathrm{GDCG@}k(v_q) =
\sum_{i=1}^{k} \frac{\mathrm{gain}(r_i)}{\log_2(i+1)}.
\]
Let $(y_1,\ldots,y_{|Y_q|})$ be the papers in $Y_q$ sorted by increasing $\mathrm{hop}(v_q,y_i)$. 
The ideal graph-weighted DCG is then defined as,
\[
\mathrm{IGDCG@}k(v_q) =
\sum_{i=1}^{\min(k,|Y_q|)}
\frac{\frac{1}{\mathrm{hop}(v_q,y_i)}}{\log_2(i+1)}.
\]
Finally, graph-weighted nDCG@$k$ is given by,
\[
\mathrm{GN@}k(v_q) =
\frac{\mathrm{GDCG@}k(v_q)}{\mathrm{IGDCG@}k(v_q)}.
\]

\subsection{Retriever}
\label{app:retriever_settings}

\textbf{Sparse Retrieval.}
BM25 indexes the \texttt{Lineage} corpus with \texttt{bm25s}\footnote{\url{https://github.com/xhluca/bm25s}} library.
Corpus and query texts are tokenized with the default BM25 scorer.
TF-IDF retrieval is implemented with \texttt{scikit-learn}\footnote{\url{https://scikit-learn.org}}. The L2-normalized corpus matrix is scored against query vectors by sparse inner products.

\textbf{Dense Retrieval.}
For dense retrievers, we evaluate Contriever, BGE-M3, Jina-Embedding-v5, Qwen3-Embedding, NV-Embedding-v2, and ReasonIR. The corresponding repository links in the HuggingFace are listed in Table~\ref{tab:huggingface}. For Qwen3-Embedding, we include the 0.6B, 4B, and 8B variants. All methods use cosine similarity as the distance metric. 

\textbf{Agentic Retrieval.}
We evaluate four agentic retrievers.
Tongyi-DeepResearch uses Tongyi-DeepResearch-30B-A3B as the search agent, with Qwen3-Embedding-4B as search backbones.
AgentIR uses \texttt{gpt-4o-mini}\footnote{\url{https://platform.openai.com/docs/models}} as the search agent and AgentIR-4B as the dense retriever. AgentIR-Qwen3-E keeps the same agent loop but replaces the retriever with Qwen3-Embedding-4B.
For PaSa, we leverage PaSa-7B-Crawler and PaSa-7B-Selector on our \textsc{Lineage}, replacing its original arXiv-based source.

\textbf{Search Engines.}
Google Search is queried through \texttt{SerpAPI}\footnote{\url{https://serpapi.com/google-search-api}}.
Queries are restricted to academic websites such as arXiv, ACL Anthology, OpenReview, and related venues, and results are limited to 2018--2025. Google Scholar retrieval also uses \texttt{SerpAPI}\footnote{\url{https://serpapi.com/google-scholar-api}} with the same year limitation.

\textbf{LLM Search.}
GPT-5.4, Gemini-3.1, and DeepSeek-V3.2 are evaluated through APIs with web search enforced. Each model receives the retrieval query, allowed venues, and the 2018--2025 year constraint, and is required to return a ranked JSON list of paper titles by the prompt in Appendix~\ref{app:llm-search-prompt}.

For corpus-based retrieval, each candidate paper is represented only by its title and abstract; full-text content is excluded from retrieval. To ensure a fair top-k evaluation, the query paper itself is removed from the retrieved candidate list, since retrievers may rank it highly based on direct lexical or semantic overlap with the query, even though it is not part of the ground-truth set.

\begin{table}[h]
\centering
\caption{Model checkpoints on HuggingFace.}
\label{tab:huggingface}
\resizebox{\linewidth}{!}{
\begin{tabular}{lll}
\toprule
\textbf{Model} & \textbf{Model checkpoints} & \textbf{License} \\
\midrule
Contriever                  & https://huggingface.co/facebook/contriever-msmarco & CC-BY-NC-4.0 \\
BGE-M3                      & https://huggingface.co/BAAI/bge-m3 & MIT \\
Jina-Embedding-v5           & https://huggingface.co/jinaai/jina-embeddings-v5-text-small & CC-BY-NC-4.0 \\
Qwen3-Embedding-0.6B        & https://huggingface.co/Qwen/Qwen3-Embedding-0.6B & Apache-2.0 \\
Qwen3-Embedding-4B          & https://huggingface.co/Qwen/Qwen3-Embedding-4B & Apache-2.0 \\
Qwen3-Embedding-8B          & https://huggingface.co/Qwen/Qwen3-Embedding-8B & Apache-2.0 \\
NV-Embedding-v2             & https://huggingface.co/nvidia/NV-Embed-v2 & CC-BY-NC-4.0 \\
ReasonIR                    & https://huggingface.co/reasonir/ReasonIR-8B & CC-BY-NC-4.0 \\
Tongyi-DeepResearch-30B-A3B & https://huggingface.co/Alibaba-NLP/Tongyi-DeepResearch-30B-A3B & Apache-2.0 \\
AgentIR-4B                  & https://huggingface.co/Tevatron/AgentIR-4B & Not specified \\
PaSa-7B-Crawler             & https://huggingface.co/bytedance-research/pasa-7b-crawler & CC-BY-NC-SA-4.0 \\
PaSa-7B-Selector            & https://huggingface.co/bytedance-research/pasa-7b-selector & CC-BY-NC-SA-4.0 \\
\bottomrule
\end{tabular}
}
\end{table}

%% file: appendix/additional-results.tex
\section{Additional Results}
\label{app:additional-result}
\subsection{Retrieval Results on \textsc{Lineage}-0.8k}
\label{app:lineage-800}
Due to API rate limits, the \textit{Search Engine} and \textit{LLM Search} methods are evaluated on a sampled subset, denoted as \textsc{Lineage}-0.8k. This subset contains 800 queries, with 100 queries evenly sampled from each query type. For a fair comparison, we also evaluate the corpus-based retrieval methods on the same subset and report the results in Table~\ref{tab:retrieval_results_800}. 
The results are consistent with the main findings in Table~\ref{tab:retrieval_results}, further supporting our conclusions drawn from the full \textsc{Lineage} dataset.

\begin{table}[t]
\centering
\caption{Evaluation results of representative retrieval methods on \textsc{Lineage}-0.8k.}
\label{tab:retrieval_results_800}
\resizebox{\linewidth}{!}{
\begin{tabular}{cl|rrrrrrrr}
\toprule
\multirow{2}{*}{\textbf{Category}} 
& \multirow{2}{*}{\textbf{Retrieval Method}} 
& \multicolumn{5}{c}{\textbf{Retrieval Metric}} 
& \multicolumn{3}{c}{\textbf{Structural Metric}} \\
\cmidrule(lr){3-7}
\cmidrule(lr){8-10}
& & \textit{R}@20 & \textit{P}@20 & \textit{H}@20 & \textit{N}@20 & \textit{M}@20 & \textit{HWR}@20 & \textit{DR}@20 & \textit{GN}@20 \\
\midrule
\multirow{2}{*}{\makecell[c]{Sparse\\Retrieval}}
& BM25
& 23.08 & 6.59 & 53.63 & 18.11 & 26.79 & 23.84 & 25.46 & 18.19 \\
& TF-IDF
& 22.17 & 6.39 & 54.00 & 17.07 & 24.63 & 22.86 & 24.43 & 17.16 \\
\midrule

\multirow{8}{*}{\makecell[c]{Dense\\Retrieval}}
& Contriever
& 19.09 & 5.80 & 50.50 & 15.09 & 22.96 & 19.71 & 20.87 & 15.00 \\
& BGE-M3 (0.6B)
& 10.84 & 2.98 & 34.00 & 8.53 & 14.25 & 11.18 & 11.85 & 8.54 \\
& Jina-Embedding-v5 (0.6B)
& 29.80 & 8.59 & 61.12 & 22.78 & 30.11 & 30.80 & 32.76 & 22.64 \\
& Qwen3-Embedding (0.6B)
& 27.70 & 8.77 & 60.50 & 22.01 & 29.80 & 28.79 & 30.79 & 22.00 \\
& Qwen3-Embedding (4B)
& \textbf{32.74} & \textbf{10.26} & 63.25 & \textbf{25.73} & 32.99 & \textbf{34.12} & \textbf{36.77} & \textbf{25.62} \\
& Qwen3-Embedding (8B)
& 30.38 & 9.80 & 61.62 & 24.39 & \underline{33.28} & 31.47 & 33.33 & 24.03 \\
& NV-Embedding-v2 (8B)
& \underline{32.02} & 9.82 & \textbf{65.38} & \underline{25.25} & \textbf{34.06} & \underline{33.35} & \underline{36.20} & \underline{25.20} \\
& ReasonIR (8B)
& 26.56 & 7.88 & 60.50 & 21.27 & 30.04 & 27.84 & 30.70 & 21.43 \\
\midrule
\multirow{4}{*}{\makecell[c]{Agentic\\Retrieval}}
& Pasa
& 22.86 & 6.66 & 54.43 & 18.98 & 27.43 & 23.45 & 24.80 & 18.71 \\
& AgentIR
& 27.61 & 8.29 & 60.12 & 22.15 & 31.31 & 28.85 & 31.43 & 22.28 \\
& AgentIR-Qwen3-Embedding
& 28.78 & 8.82 & 59.25 & 22.93 & 30.98 & 29.88 & 31.81 & 22.82 \\
& Tongyi-DeepResearch
& 31.87 & \underline{9.92} & \underline{63.88} & 24.99 & 33.09 & 33.23 & 35.89 & 24.99 \\

\midrule

\multirow{2}{*}{\makecell[c]{Search\\Engine}}
& Google Search
& 9.64 & 0.40 & 32.37 & 6.64 & 10.62 & 9.99 & 2.23 & 1.42 \\
& Google Scholar
& 7.57 & 2.00 & 23.62 & 5.23 & 8.11 & 8.01 & 8.78 & 5.14 \\
\midrule

\multirow{3}{*}{\makecell[c]{LLM\\Search}}
& GPT-5.4 
& 7.24 & 3.30 & 29.13 & 7.12 & 13.56 & 7.42 & 7.87 & 6.65 \\
& Gemini-3.1 
& 4.47 & 1.97 & 21.50 & 4.58 & 10.35 & 4.51 & 4.46 & 4.23 \\
& DeepSeek-V3.2 
& 4.52 & 1.82 & 18.12 & 4.55 & 8.92 & 4.51 & 4.59 & 4.23 \\
\bottomrule
\end{tabular}}
\end{table}

\subsection{Method-Level Results across Lineage Hops}
\label{app:hop_results}
Table~\ref{tab:hop_recall_results} presents Recall@20 for each representative retrieval method across one-hop to seven-hop lineage targets. The method-level results reveal a consistent trend across hops that Recall@20 generally decreases as hop distance increases, showing that longer-hop empirical lineage papers are harder to retrieve than direct empirical comparison neighbors. For the direct one-hop setting, Tongyi-DeepResearch achieves the best Recall@20, followed by Qwen3-E. (4B). When averaging uniformly over all hop columns, Qwen3-E. (8B) performs best, followed by Qwen3-E. (4B).

\begin{table}[t]
\centering
\caption{Recall@20 results of retrieval methods on \textsc{Lineage} across hops.}
\label{tab:hop_recall_results}
\footnotesize
\resizebox{\linewidth}{!}{\begin{tabular}{cl|rrrrrrr}
\toprule
\textbf{Category} & \textbf{Retrieval Method} & \textbf{1-\textit{hop}} & \textbf{2-\textit{hop}} & \textbf{3-\textit{hop}} & \textbf{4-\textit{hop}} & \textbf{5-\textit{hop}} & \textbf{6-\textit{hop}} & \textbf{7-\textit{hop}} \\
\midrule
\rowcolor{gray!20} \multicolumn{2}{l|}{\textit{w/} \textsc{Lineage} \textit{Corpus} (Avg.)}
& 28.42 & 14.02 & 6.68 & 3.40 & 1.74 & 0.69 & 0.83 \\
\midrule
\multirow{2}{*}{\makecell[c]{Sparse\\Retrieval}}
& BM25
& 25.74 & 11.93 & 5.43 & 2.23 & 1.26 & 2.08 & 1.79 \\
& TF-IDF
& 23.64 & 11.53 & 5.71 & 2.61 & 1.43 & 1.14 & 0.00 \\
\midrule
\multirow{8}{*}{\makecell[c]{Dense\\Retrieval}}
& Contriever
& 21.19 & 9.88 & 4.67 & 2.11 & 1.08 & 0.00 & 0.00 \\
& BGE-M3 (0.6B)
& 12.34 & 5.98 & 3.41 & 1.74 & 0.88 & 0.00 & 2.68 \\
& Jina-Embedding-v5 (0.6B)
& 30.46 & 14.60 & 6.59 & 2.97 & 2.29 & 0.38 & 0.00 \\
& Qwen3-Embedding (0.6B)
& 30.18 & 15.04 & 7.13 & 3.50 & 2.27 & 0.00 & 1.79 \\
& Qwen3-Embedding (4B)
& 35.09 & 18.75 & 9.16 & 4.24 & 2.98 & 0.13 & 0.00 \\
& Qwen3-Embedding (8B)
& 33.18 & 18.79 & 9.76 & 5.16 & 4.18 & 2.45 & 0.00 \\
& NV-Embedding-v2 (8B)
& 34.38 & 17.44 & 7.64 & 3.03 & 2.31 & 0.18 & 1.79 \\
& ReasonIR (8B)
& 29.11 & 14.91 & 7.99 & 3.44 & 3.12 & 1.96 & 3.57 \\
\midrule
\multirow{4}{*}{\makecell[c]{Agentic\\Retrieval}}
& Pasa
& 23.47 & 11.86 & 6.06 & 2.76 & 1.45 & 1.30 & 0.00 \\
& AgentIR$^\diamond$
& 31.43 & 13.13 & 6.01 & 4.32 & 0.00 & 0.00 & 0.00 \\
& AgentIR-Qwen3-Embedding$^\diamond$
& 31.81 & 15.47 & 6.00 & 4.15 & 0.00 & 0.00 & 0.00 \\
& Tongyi-DeepResearch$^\diamond$
& 35.89 & 16.94 & 7.93 & 5.37 & 1.04 & 0.00 & 0.00 \\
\midrule
\rowcolor{gray!20} \multicolumn{2}{l|}{\textit{w/} \textit{Online Sources} (Avg.)}
& 6.48 & 4.61 & 2.16 & 1.93 & 0.05 & 0.00 & 0.00 \\
\midrule
\multirow{2}{*}{\makecell[c]{Search\\Engine}}
& Google Search$^\diamond$
& 10.54 & 3.97 & 0.91 & 2.64 & 0.24 & 0.00 & 0.00 \\
& Google Scholar$^\diamond$
& 8.90 & 2.38 & 1.75 & 2.23 & 0.00 & 0.00 & 0.00 \\
\midrule
\multirow{3}{*}{\makecell[c]{LLM\\Search}}
& GPT-5.4$^\diamond$
& 7.24 & 7.87 & 9.61 & 5.40 & 2.27 & 0.00 & 0.00 \\
& Gemini-3.1$^\diamond$
& 4.47 & 4.46 & 5.08 & 2.74 & 2.04 & 0.00 & 0.00 \\
& DeepSeek-V3.2$^\diamond$
& 4.52 & 4.59 & 6.72 & 1.85 & 0.45 & 0.00 & 0.00 \\
\bottomrule
\end{tabular}}
\end{table}

\subsection{Method-Level Results across Query Types}
\label{app:query_type_results}

Tables~\ref{tab:query_type_recall_results} and~\ref{tab:query_type_hwr_results} report Recall@20 and HWR@20 for each representative retrieval method across the eight query types.
The query-type columns use the following abbreviations.
\bi{Title} denotes the direct title query, and \bi{Title-Ctx} denotes the title-context query formed by concatenating the title and abstract.
\bi{Top-KW} and \bi{Top-Sem} refer to topical keyword and topical semantic queries, respectively.
\bi{Ctx-KW} and \bi{Ctx-Sem} denote context keyword and context semantic queries focused on task motivation and application domain, while \bi{Meth-KW} and \bi{Meth-Sem} denote method keyword and method semantic queries focused on method family, modeling idea, or technical mechanism.

The method-level results reveal the same pattern as illustrated in Table~\ref{tab:retrieval_query_results}. Different retrieval families perform better on different query types, with agentic methods being strong on title, topical semantic, and context keyword queries, while dense retrievers remain competitive on topical keyword, context semantic, and method semantic queries. Across query types, Tongyi-DeepResearch achieves the best average Recall@20 and HWR@20, followed by Qwen3-E. (4B), while AgentIR-Qwen3-E performs best on title-context queries and NV-E.-v2 performs best on context-semantic queries.

\begin{table*}[t]
\centering
\caption{Recall@20 results of retrieval methods on \textsc{Lineage} across query types. For each method, the best query type is highlighted in \textbf{bold}, and the second-best is \underline{underlined}.}
\label{tab:query_type_recall_results}
\resizebox{\linewidth}{!}{
\begin{tabular}{cl|rrrrrrrr}
\toprule
\multirow{2}{*}{\textbf{Category}}
& \multirow{2}{*}{\textbf{Retrieval Method}}
& \multicolumn{8}{c}{\textbf{Query Type}} \\
\cmidrule(lr){3-10}
& & Title & Title-Ctx & Top-KW & Top-Sem & Ctx-KW & Ctx-Sem & Meth-KW & Meth-Sem \\
\midrule
\rowcolor{gray!20} \multicolumn{2}{l|}{\textit{w/} \textsc{Lineage} \textit{Corpus} (Avg.)}
& 30.36 & 39.04 & 29.19 & 31.07 & 18.90 & 26.07 & 7.46 & 17.14 \\
\midrule
\multirow{2}{*}{\makecell[c]{Sparse\\Retrieval}}
& BM25
& \underline{27.15} & \textbf{36.85} & 26.10 & 26.58 & 16.20 & 21.39 & 7.34 & 14.88 \\
& TF-IDF
& 24.05 & \textbf{36.31} & 23.21 & \underline{24.25} & 14.85 & 20.10 & 6.72 & 13.86 \\
\midrule
\multirow{8}{*}{\makecell[c]{Dense\\Retrieval}}
& Contriever
& \underline{22.17} & \textbf{31.71} & 21.68 & 21.92 & 13.09 & 19.09 & 4.96 & 11.00 \\
& BGE-M3 (0.6B)
& 10.68 & \textbf{28.14} & \underline{11.72} & 10.69 & 6.98 & 9.30 & 2.43 & 5.72 \\
& Jina-Embedding-v5 (0.6B)
& 31.97 & \textbf{42.08} & 30.40 & \underline{32.39} & 19.58 & 27.94 & 7.48 & 18.23 \\
& Qwen3-Embedding (0.6B)
& 31.68 & \textbf{40.34} & 30.09 & \underline{32.80} & 19.48 & 27.32 & 7.29 & 18.67 \\
& Qwen3-Embedding (4B)
& 37.02 & \textbf{45.98} & 35.38 & \underline{38.12} & 23.59 & 32.07 & 9.92 & 23.03 \\
& Qwen3-Embedding (8B)
& 35.28 & \textbf{45.41} & 33.46 & \underline{36.21} & 22.01 & 29.75 & 9.82 & 22.87 \\
& NV-Embedding-v2 (8B)
& 36.76 & \textbf{41.42} & 34.86 & \underline{37.73} & 23.05 & 31.33 & 9.55 & 23.37 \\
& ReasonIR (8B)
& 29.96 & \textbf{42.60} & 27.94 & \underline{32.13} & 17.69 & 25.20 & 7.94 & 19.55 \\
\midrule
\multirow{4}{*}{\makecell[c]{Agentic\\Retrieval}}
& Pasa
& 26.81 & \textbf{30.22} & 25.29 & \underline{27.66} & 16.60 & 23.47 & 7.24 & 15.76 \\
& AgentIR$^\diamond$
& 33.63 & \textbf{40.24} & 34.07 & \underline{37.52} & 20.99 & 31.52 & 6.63 & 16.31 \\
& AgentIR-Qwen3-Embedding$^\diamond$
& \underline{37.42} & \textbf{38.53} & 36.87 & 35.11 & 25.23 & 32.67 & 7.71 & 16.69 \\
& Tongyi-DeepResearch$^\diamond$
& 40.43 & \textbf{46.69} & 37.62 & \underline{41.84} & 25.19 & 33.89 & 9.34 & 19.99 \\
\midrule
\rowcolor{gray!20} \multicolumn{2}{l|}{\textit{w/} \textit{Online Sources} (Avg.)}
& 6.06 & 3.76 & 8.96 & 7.56 & 5.85 & 6.50 & 3.06 & 5.02 \\
\midrule
\multirow{2}{*}{\makecell[c]{Search\\Engine}}
& Google Search$^\diamond$
& \underline{15.51} & 2.04 & \textbf{17.83} & 14.04 & 6.37 & 12.94 & 2.17 & 6.19 \\
& Google Scholar$^\diamond$
& 1.02 & 0.00 & \textbf{14.80} & \underline{12.78} & 10.52 & 7.00 & 5.84 & 8.57 \\
\midrule
\multirow{3}{*}{\makecell[c]{LLM\\Search}}
& GPT-5.4$^\diamond$
& 7.55 & \textbf{8.63} & \underline{8.05} & 6.94 & 6.76 & 6.81 & 5.21 & 7.97 \\
& Gemini-3.1$^\diamond$
& \underline{6.20} & \textbf{6.85} & 5.53 & 3.92 & 4.91 & 4.08 & 1.89 & 2.35 \\
& DeepSeek-V3.2$^\diamond$
& \underline{6.44} & \textbf{6.89} & 4.87 & 4.08 & 4.87 & 4.27 & 2.06 & 2.67 \\
\bottomrule
\end{tabular}}
\end{table*}

\begin{table*}[t]
\centering
\caption{HWR@20 results of representative retrieval methods on \textsc{Lineage} across query types.}
\label{tab:query_type_hwr_results}
\resizebox{\linewidth}{!}{
\begin{tabular}{cl|rrrrrrrr}
\toprule
\multirow{2}{*}{\textbf{Category}}
& \multirow{2}{*}{\textbf{Retrieval Method}}
& \multicolumn{8}{c}{\textbf{Query Type}} \\
\cmidrule(lr){3-10}
& & Title & Title-Ctx & Top-KW & Top-Sem & Ctx-KW & Ctx-Sem & Meth-KW & Meth-Sem \\
\midrule
\rowcolor{gray!20} \multicolumn{2}{l|}{\textit{w/} \textsc{Lineage} \textit{Corpus} (Avg.)}
& 31.89 & \textbf{41.19} & 30.63 & \underline{32.67} & 19.57 & 27.06 & 7.80 & 18.01 \\
\midrule
\multirow{2}{*}{\makecell[c]{Sparse\\Retrieval}}
& BM25
& \underline{28.66} & \textbf{39.11} & 27.51 & 28.09 & 16.94 & 22.42 & 7.75 & 15.74 \\
& TF-IDF
& 25.27 & \textbf{38.47} & 24.39 & \underline{25.56} & 15.44 & 21.00 & 7.08 & 14.61 \\
\midrule
\multirow{8}{*}{\makecell[c]{Dense\\Retrieval}}
& Contriever
& \underline{23.45} & \textbf{33.60} & 22.84 & 23.19 & 13.59 & 19.91 & 5.23 & 11.64 \\
& BGE-M3 (0.6B)
& 11.24 & \textbf{29.82} & \underline{12.31} & 11.22 & 7.20 & 9.58 & 2.55 & 5.98 \\
& Jina-Embedding-v5 (0.6B)
& 33.73 & \textbf{44.52} & 32.05 & \underline{34.21} & 20.38 & 29.17 & 7.89 & 19.20 \\
& Qwen3-Embedding (0.6B)
& 33.42 & \textbf{42.65} & 31.72 & \underline{34.66} & 20.36 & 28.59 & 7.69 & 19.64 \\
& Qwen3-Embedding (4B)
& 38.81 & \textbf{48.45} & 37.11 & \underline{40.03} & 24.52 & 33.43 & 10.37 & 24.11 \\
& Qwen3-Embedding (8B)
& 36.80 & \textbf{47.74} & 34.92 & \underline{37.82} & 22.73 & 30.80 & 10.19 & 23.81 \\
& NV-Embedding-v2 (8B)
& 38.78 & \textbf{43.76} & 36.69 & \underline{39.78} & 23.98 & 32.67 & 10.01 & 24.60 \\
& ReasonIR (8B)
& 31.44 & \textbf{45.20} & 29.24 & \underline{33.77} & 18.24 & 26.07 & 8.30 & 20.56 \\
\midrule
\multirow{4}{*}{\makecell[c]{Agentic\\Retrieval}}
& Pasa
& 28.22 & \textbf{31.88} & 26.57 & \underline{29.13} & 17.27 & 24.49 & 7.62 & 16.59 \\
& AgentIR$^\diamond$
& 35.20 & \textbf{42.59} & 35.45 & \underline{39.31} & 21.74 & 32.72 & 6.82 & 16.95 \\
& AgentIR-Qwen3-Embedding$^\diamond$
& \underline{38.98} & \textbf{40.11} & 38.51 & 36.64 & 25.91 & 33.47 & 7.88 & 17.55 \\
& Tongyi-DeepResearch$^\diamond$
& 42.40 & \textbf{48.78} & 39.46 & \underline{44.01} & 25.67 & 34.58 & 9.83 & 21.11 \\
\midrule
\rowcolor{gray!20} \multicolumn{2}{l|}{\textit{w/} \textit{Online Sources} (Avg.)}
& 6.27 & 3.94 & \textbf{9.40} & \underline{8.00} & 6.02 & 6.54 & 3.16 & 5.28 \\
\midrule
\multirow{2}{*}{\makecell[c]{Search\\Engine}}
& Google Search$^\diamond$
& \underline{15.98} & 2.27 & \textbf{18.76} & 14.47 & 6.61 & 12.91 & 2.34 & 6.60 \\
& Google Scholar$^\diamond$
& 1.04 & 0.00 & \textbf{15.95} & \underline{13.85} & 10.68 & 7.18 & 6.11 & 9.24 \\
\midrule
\multirow{3}{*}{\makecell[c]{LLM\\Search}}
& GPT-5.4$^\diamond$
& 7.80 & \textbf{9.04} & 8.09 & 7.37 & 6.76 & 6.83 & 5.30 & \underline{8.18} \\
& Gemini-3.1$^\diamond$
& \underline{6.33} & \textbf{6.94} & 5.49 & 4.00 & 5.06 & 4.02 & 1.82 & 2.42 \\
& DeepSeek-V3.2$^\diamond$
& \underline{6.56} & \textbf{6.87} & 4.82 & 4.11 & 5.00 & 4.21 & 1.92 & 2.63 \\
\bottomrule
\end{tabular}}
\end{table*}

%% file: appendix/lineage-ha.tex
\section{\textsc{Lineage}-HA}
\label{app:lineage-ha}

\begin{figure}[tb]
    \centering
    \includegraphics[width=\linewidth]{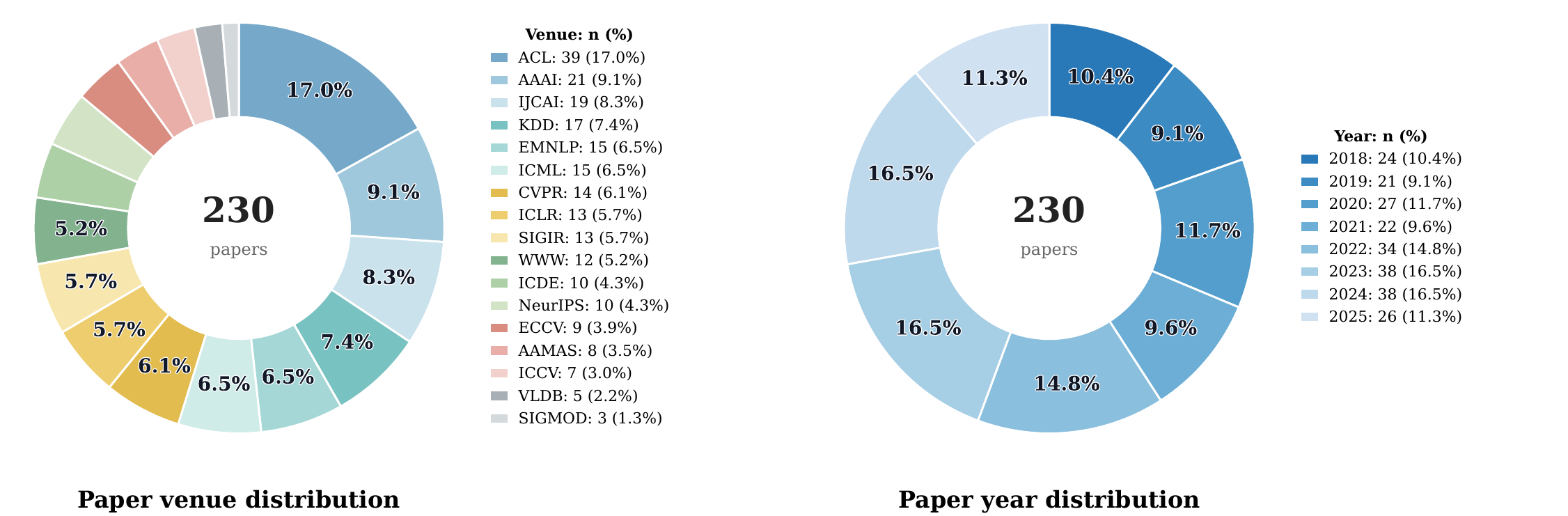}
    \caption{The venue and year distribution of the 230 papers annotated for paper type in \textsc{Lineage}-HA.}
    \label{fig:ha-pie}
    \vspace{-1em}
\end{figure}

To validate the proposed paper type classifier and dual-source lineage extractor, we construct a human-annotated validation set termed \textsc{Lineage}-HA. It contains paper-type annotations for 230 papers and lineage-relation annotations for 150 methodology papers. The lineage-relation annotations contain 970 positive lineage relations, with an average of 6.47 relations per paper. The annotated papers cover different publication years, venues, and research areas, as illustrated in Figure~\ref{fig:ha-pie}.

\subsection{Human Annotation Protocol}
\label{sec:ha-annotation}

The annotations were produced by recruited independent annotators who are PhD students working in research areas related to the annotated papers. Each annotator was compensated at a rate of USD 10 per hour.

For paper-type annotation, annotators assign each paper to its primary contribution. A paper is labeled as \textit{methodology} if its main contribution is a technical method, as \textit{benchmark} if it mainly contributes an evaluation resource or protocol, and as \textit{survey} if it mainly synthesizes prior work.

For lineage-relation annotation, annotators inspect the full paper, including experimental text, tables, figures, captions, and appendices. They identify methods explicitly used as empirical comparison baselines, map method names and aliases to bibliography entries, and exclude citations that are not used for empirical comparison. Each paper is independently annotated by two annotators. Disagreements and uncertain cases are reviewed before determining the final annotations.

\subsection{Annotation Agreement}
\label{sec:annotation-agreement}

To assess the reliability of the lineage-relation annotations, we conduct an agreement analysis on a 50-paper subset of \textsc{Lineage}-HA. One author independently annotates the same papers using the annotation protocol in Section~\ref{sec:ha-annotation}, without access to the labels produced by the independent annotators. The author annotations are used only to assess annotation quality and are not used to construct the final annotations or evaluate the lineage extractor.

For each paper, we compare the sets of lineage relations produced by the two independent annotators and the author. We use Krippendorff's alpha with MASI distance to measure agreement among the three set-valued annotations. We additionally report pairwise micro-averaged S{\o}rensen--Dice scores. Agreement is computed only from the three independently produced annotation sets, while the adjudicated final annotations are excluded.

\begin{table}[t]
\centering
\caption{Annotation agreement on a 50-paper subset of \textsc{Lineage}-HA.}
\label{tab:annotation-agreement}
\small
\begin{tabular}{llc}
\toprule
\textbf{Metric} & \textbf{Annotation Sets} & \textbf{Score} \\
\midrule
Krippendorff's $\alpha$
& Annotator 1, Annotator 2, Author & 0.8490 \\
Micro Dice
& Annotator 1--Annotator 2 & 0.9757 \\
Micro Dice
& Annotator 1--Author & 0.9706 \\
Micro Dice
& Annotator 2--Author & 0.9805 \\
\bottomrule
\end{tabular}
\end{table}

As shown in Table~\ref{tab:annotation-agreement}, Krippendorff's alpha reaches 0.8490, while all pairwise Micro-Dice scores exceed 0.97. These results indicate that the annotations produced by the independent annotators are highly consistent with each other and with annotations independently produced by an author under the same evidence-based protocol.

\subsection{Paper Type Classification Task}
\label{sec:paper-type-classification}

We sample 230 papers across venues and publication years for paper-type annotation, including 150 methodology papers, 40 benchmark papers, and 40 survey papers, as illustrated in Figure~\ref{fig:ha-pie}. Based on this annotated set, we evaluate our paper type classifier against several alternatives, including:

\begin{itemize}
    \item \textbf{Keyword Classifier.}
    This classifier serves as a lightweight heuristic baseline, relying on keyword matching to distinguish survey, benchmark, and methodology papers.

    \item \textbf{LLM \textit{w}/ Title \& Abstract (Used in our pipeline).}
    This method uses an LLM to infer the primary contribution type from the paper title and abstract, balancing semantic understanding with minimal input cost.

    \item \textbf{LLM \textit{w}/ Title \& Abstract \& Introduction.}
    This variant provides the LLM with the introduction in addition to the title and abstract, enabling more informed classification.
\end{itemize}

\begin{figure}[t]
    \centering
    \includegraphics[width=\linewidth]{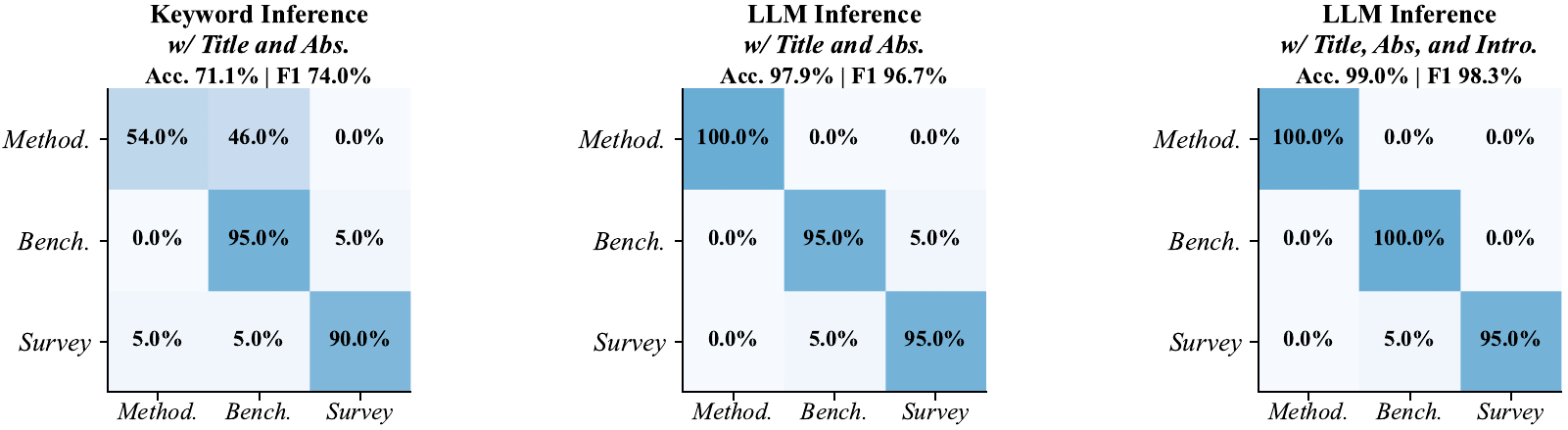}
    \caption{Comparison of paper-type inference strategies on the 230-paper paper-type annotation set in \textsc{Lineage}-HA. Each confusion matrix reports predictions over methodology, benchmark, and survey papers, with overall accuracy and macro-F1 shown above each matrix.}
    \label{fig:paper-type-infe}
\end{figure}

Figure~\ref{fig:paper-type-infe} compares three paper-type inference strategies on the 230-paper paper-type annotation set in \textsc{Lineage}-HA. The keyword-based classifier performs substantially worse than the LLM-based classifiers, particularly on methodology papers. In contrast, the LLM classifier using only titles and abstracts achieves strong performance across methodology, benchmark, and survey papers. Adding the introduction provides only a marginal improvement while increasing input length and parsing cost. We therefore use the title-and-abstract LLM classifier in our \textsc{Lineage} construction pipeline.

\subsection{Lineage Relation Extraction Task}
\label{sec:lineage-relation-extraction}

The lineage-relation annotation set contains 150 methodology papers and 970 positive lineage relations, corresponding to an average of 6.47 relations per paper. We evaluate our dual-source lineage extractor and its alternatives on \textsc{Lineage}-HA.

The extraction methods include:
\begin{itemize}
    \item \textbf{Extractor \textit{w}/ Integration (Used in our pipeline).}
    This method performs extraction over both MinerU- and GROBID-parsed full texts and integrates their outputs through an LLM-based resolution step. The dual-source integration leverages complementary parsing results to improve coverage and robustness against parser-specific errors.

    \item \textbf{Extractor \textit{w}/ Verifier.}
    These variants first extract candidate lineage relations and then ask the LLM to verify them against the paper content, aiming to reduce unsupported baseline predictions.

    \item \textbf{Extractor \textit{w}/ Filtering.}
    These variants introduce an additional filtering stage to locate likely experimental comparison sections before performing lineage extraction.

    \item \textbf{Extractor \textit{w}/ Evidence.}
    These methods extract lineage relations from selected evidence units and reference entries, focusing the LLM on comparison-relevant content rather than the entire paper.
\end{itemize}

Table~\ref{tab:lineage_extraction_source} reports the lineage extraction performance of different extractor variants with MinerU, GROBID, and their dual-source combination. Among all variants, the dual-source extractor with integration achieves the best micro-F1 score of 92.95, with a recall of 93.71 and the highest precision of 92.21. Although the dual-source verifier obtains the highest recall of 95.50, its lower precision indicates that additional verification does not sufficiently suppress unsupported lineage candidates. The filtering and evidence-based variants perform worse overall, suggesting that empirical baseline evidence is not always confined to a small set of pre-selected sections or evidence units.

The comparison across input sources further shows that MinerU and GROBID provide complementary paper representations. For the integration-based extractor, MinerU alone achieves high recall but substantially lower precision, while GROBID alone yields lower overall performance. Combining the two parsing sources improves precision while preserving high recall, resulting in the strongest overall micro-F1. We therefore adopt the dual-source extractor with integration in the \textsc{Lineage}-Constructor pipeline.

\begin{table}[t]
\centering
\caption{Lineage extraction results of our dual-source lineage extractor and its variants on \textsc{Lineage}-HA with different input sources.}
\label{tab:lineage_extraction_source}
\small
\begin{tabular}{@{\hspace{1em}}ll@{\hspace{1em}}|@{\hspace{2em}}c@{\hspace{2em}}c@{\hspace{2em}}c@{\hspace{1em}}}
\toprule
\textbf{Extractor} & \textbf{Source}
& \textbf{Recall} & \textbf{Precision} & \textbf{micro F1} \\
\midrule

\multirow[c]{3}{*}{\textit{w}/ Integration}
& MinerU & \underline{94.24} & 83.71 & 88.66 \\
& GROBID & 85.25 & 81.87 & 83.52 \\
& Dual & 93.71 & \textbf{92.21} & \textbf{92.95} \\
\midrule

\multirow[c]{3}{*}{\textit{w}/ Verifier}
& MinerU & 93.53 & \underline{92.04} & \underline{92.77} \\
& GROBID & 80.22 & 86.27 & 83.13 \\
& Dual & \textbf{95.50} & 84.15 & 89.47 \\
\midrule

\multirow[c]{3}{*}{\textit{w}/ Filtering}
& MinerU & 75.18 & 88.00 & 81.09 \\
& GROBID & 65.83 & 76.09 & 70.59 \\
& Dual & 69.99 & 86.62 & 77.26 \\
\midrule

\multirow[c]{3}{*}{\textit{w}/ Evidence}
& MinerU & 85.79 & 87.85 & 86.81 \\
& GROBID & 67.99 & 78.10 & 72.69 \\
& Dual & 85.79 & 89.66 & 87.68 \\
\bottomrule
\end{tabular}
\end{table}

%% file: appendix/paper-collection.tex
\section{Construction Details}
\subsection{Metadata and PDF Acquisition}
\label{app:pdf-acquistion}
\textbf{Acquisition Sources.} We acquire paper PDFs through a venue-aware collection pipeline that integrates 11 open-access sources. Starting from a canonical venue-year paper list, the collector queries venue-specific PDF sources in a fixed priority order, as shown in Table~\ref{tab:collection_sources}. Official proceedings are used as the primary source when available, while OpenReview and arXiv serve as complementary fallback sources for venues with incomplete official coverage or restricted automated access. After a PDF is obtained, the file is validated before downstream parsing, and abnormal or incomplete files are filtered to reduce noise in lineage extraction.

\begin{table*}[h]
\centering
\caption{Venue-specific sources for metadata supplementation and PDF acquisition. PDF sources are acquired from left to right.}
\label{tab:collection_sources}
\begin{tabular}{l|l|l}
\toprule
\textbf{Venue} & \textbf{PDF Acquisition Order} & \textbf{Official OA Source} \\
\midrule
AAAI    & AAAI OJS $\rightarrow$ arXiv & \href{https://ojs.aaai.org/index.php/AAAI}{AAAI OJS} \\
AAMAS \qquad ~   & IFAAMAS $\rightarrow$ OpenReview $\rightarrow$ arXiv \qquad ~ & \href{https://www.ifaamas.org/proceedings.html}{IFAAMAS} \\
ACL     & ACL Anthology $\rightarrow$ arXiv & \href{https://aclanthology.org/}{ACL Anthology} \\
ECCV    & ECVA $\rightarrow$ arXiv & \href{https://www.ecva.net/papers.php}{ECVA} \\
CVPR    & CVF $\rightarrow$ OpenReview $\rightarrow$ arXiv & \href{https://openaccess.thecvf.com/}{CVF} \\
EMNLP   & ACL Anthology $\rightarrow$ arXiv & \href{https://aclanthology.org/}{ACL Anthology} \\
ICCV    & CVF $\rightarrow$ arXiv & \href{https://openaccess.thecvf.com/}{CVF} \\
ICDE    & arXiv & -- \\
ICLR    & OpenReview $\rightarrow$ arXiv & \href{https://openreview.net/}{OpenReview} \\
ICML    & PMLR $\rightarrow$ OpenReview $\rightarrow$ arXiv & \href{https://proceedings.mlr.press/}{PMLR} \\
IJCAI   & IJCAI $\rightarrow$ arXiv & \href{https://www.ijcai.org/proceedings/}{IJCAI} \\
KDD$^\dagger$     & OpenReview $\rightarrow$ arXiv & -- \\
NeurIPS & NeurIPS $\rightarrow$ OpenReview $\rightarrow$ arXiv & \href{https://proceedings.neurips.cc/}{NeurIPS} \\
SIGIR$^\dagger$   & arXiv & -- \\
SIGMOD$^\dagger$  & arXiv & -- \\
VLDB    & PVLDB $\rightarrow$ arXiv & \href{https://www.vldb.org/pvldb/}{PVLDB} \\
WWW$^\dagger$     & OpenReview $\rightarrow$ arXiv & -- \\
\bottomrule
\multicolumn{3}{p{0.9\textwidth}}{\footnotesize
$^\dagger$ACM proceedings are openly accessible through the ACM Digital Library, but it does not provide a public bulk-download API and restricts programmatic access. We therefore do not use it as an automated PDF source.} \\
\end{tabular}
\end{table*}

\subsection{Quality Filtering and Calibration}
\label{app:metadata-quality}
\paragraph{PDF Filtering.} After metadata enrichment and PDF validation, we apply two lightweight diagnostic filters to reduce noisy records before downstream parsing and lineage extraction. First, we remove PDFs with fewer than six pages. This threshold serves as a conservative sanity check, since regular research papers in the venues covered by \textsc{Lineage} are typically longer, whereas shorter files are more likely to be extended abstract-like non-regular-track records collected from heterogeneous open-access sources. It reduces the possibility that incomplete or irrelevant documents enter the full-paper extraction stage.

\paragraph{Abstract Calibration.} We inspect abstracts with abnormal surface forms, including abstracts shorter than 200 characters, longer than 2000 characters, or lacking terminal punctuation. These cases are likely to reflect metadata inconsistencies rather than genuine abstract content, such as truncated abstracts, duplicated boilerplate text, or incorrectly captured introduction paragraphs. We manually review and correct these records to maintain a consistent metadata representation for query generation and lineage construction.

\paragraph{Graph Calibration.}
Table~\ref{tab:calibration-diagnostics} reports the stage-wise effect of lineage graph calibration, where we employ two local graph diagnostics to demonstrate the properties of graph structures. The clustering coefficient~\cite{watts1998collective} measures local neighborhood closure, namely how often nodes sharing a neighbor are also connected. The edge-neighborhood Jaccard similarity~\cite{onnela2007structure} measures the overlap between the neighbor sets of two nodes connected by a lineage edge.

Temporal calibration removes 436 future-year and cycle-forming edges, accounting for 0.15\% of the raw matched edges.
It mainly enforces the temporal semantics of lineage relations, where baseline papers should not appear after their corresponding source papers. Structural calibration then corrects abnormal connectivity caused by widely reused baseline methods, which can link otherwise distinct method families and make multi-hop empirical comparison neighborhoods overly broad.
It restricts upper-tail connectivity and yields a calibrated graph with 67,502 papers and 164,948 lineage edges.
The clustering coefficient increases to 0.3035, and the mean edge-neighborhood Jaccard similarity increases to 0.1420, suggesting stronger local structural coherence after calibration.

\begin{table}[t]
\centering
\caption{Stage-wise effect of lineage graph calibration.}
\label{tab:calibration-diagnostics}
\small
\begin{tabular}{lrrrr}
\toprule
\textbf{Stage} & \textbf{Node} & \textbf{Edge}  
& \textbf{Clustering} & \textbf{Edge Jaccard}\\
\midrule
Raw Lineage Graph 
& 89,485 & 238,511 & 0.1627 & 0.1022 \\

Temporal Calibration 
& 89,441 & 238,075 &  0.1635 & 0.1020 \\

Structural Calibration 
& 67,502 & 164,948 &  0.3035 & 0.1420 \\
\bottomrule
\end{tabular}
\end{table}

\subsection{Cost Comparison}
\label{app:cost}
We estimate the annotation cost of our automatic labeling system based on two tasks. In our \textsc{Lineage}-HA labeling, paper type classification takes approximately 20--60 seconds per paper. For 121,048 papers, this requires around 672--2,017 hours. Lineage relation extraction takes approximately 3-10 minutes per paper. For 112,996 papers, this requires around 5,650--18,833 hours. With a compensation rate of 10 USD per hour, manually completing both tasks would cost around 63.2K--208.5K USD.

In comparison, our automatic system takes only 0.8--1.2 seconds per paper for paper type classification and 15--28 seconds per paper for lineage relation extraction. This corresponds to approximately 497.7--919.2 GPU hours on a single Nvidia H20 GPU. Accounting for GPU depreciation and electricity, the total computational cost is around 228--491 USD. Even using cloud GPU services for estimation, the cost remains below 0.7K USD. This indicates that the automatic pipeline reduces annotation cost by roughly two orders of magnitude.

%% file: appendix/statistics.tex
\section{\textsc{Lineage} Statistics}
\label{app:lineage-stat}

In this section, we provide additional statistics for \textsc{Lineage}, complementing the corpus-scale summary, lineage graph analysis, and query distribution.

\subsection{Corpus Coverage and Source Composition}
\label{app:corpus-coverage}

\begin{figure}[h]
    \centering
    \includegraphics[width=1\linewidth]{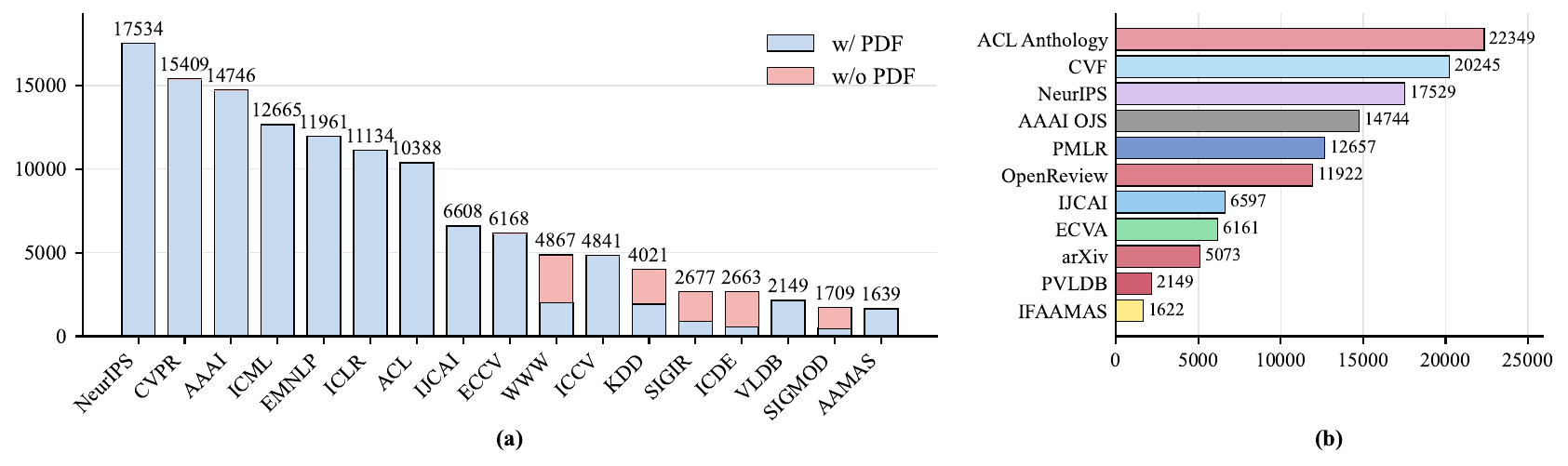}
    \caption{PDF acquisition coverage and source composition. \textbf{(a)} Venue-level distribution of papers with and without open-access PDFs. \textbf{(b)} Source-level distribution of successfully acquired PDFs.}
    \label{fig:metadata-pdf}
\end{figure}

Figure~\ref{fig:metadata-pdf}\textcolor{red}{(a)} reports the venue-level PDF acquisition coverage. Across 131,179 collected paper records, 121,048 papers have valid PDFs, while 10,131 papers remain without PDFs, giving an overall PDF coverage of 92.3\%. The largest venues, including NeurIPS, CVPR, AAAI, and ICML, contribute most of the collected papers and also account for a large portion of the acquired PDFs. Figure~\ref{fig:metadata-pdf}\textcolor{red}{(b)} summarizes the successful PDF sources, with the source-level counts summing to 121,048. ACL Anthology, CVF, NeurIPS, AAAI OJS, PMLR, and OpenReview are the main contributors, while arXiv and smaller venue-specific sources recover additional papers that are not covered by the primary sources.

\subsection{Lineage Graph Structure}
\label{app:lineage-structure}

\begin{figure}[t]
    \centering
    \includegraphics[width=1\linewidth]{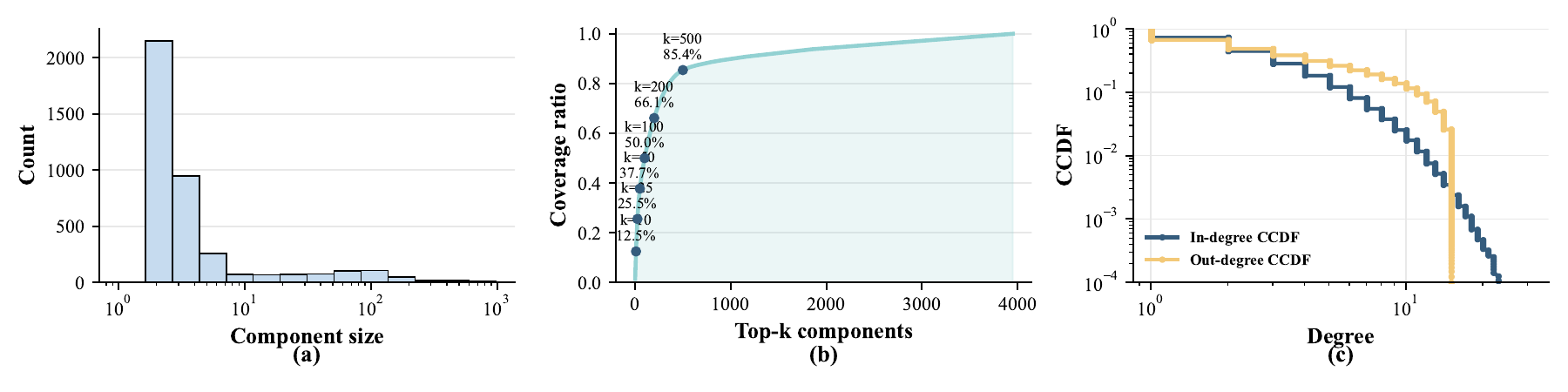}
    \caption{Structural properties of the constructed lineage graphs. \textbf{(a)} Distribution of component sizes. \textbf{(b)} Node coverage by the largest components. \textbf{(c)} In-degree and out-degree distributions.}
    \label{fig:structure-analysis}
\end{figure}

We further analyze the structural properties of the constructed lineage graphs. Figures~\ref{fig:structure-analysis}\textcolor{red}{(a)} and \textcolor{red}{(b)} show that \textsc{Lineage} contains many small method families together with a smaller number of large components. This pattern is expected in AI literature, where most methods are compared within relatively local research areas, while widely used methods can connect broader methodological regions. The degree distribution in Figure~\ref{fig:structure-analysis}\textcolor{red}{(c)} also exhibits a long-tailed pattern, suggesting that a small number of papers serve as common empirical baselines or later comparison anchors, whereas most papers have a limited number of direct lineage neighbors. 

\begin{wrapfigure}{r}{0.42\linewidth}
    \centering
    \vspace{-1em}
    \includegraphics[width=0.9\linewidth]{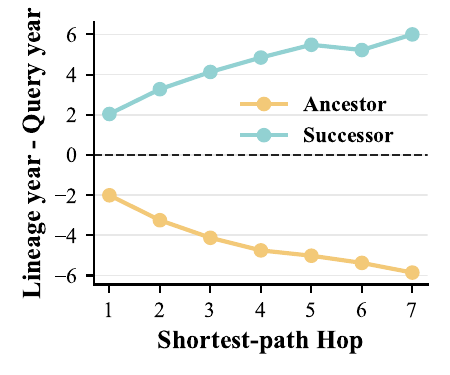}
    \caption{Average publication-year offsets over lineage distance.}
    \label{fig:year-gap-by-hop}
    \vspace{-1em}
\end{wrapfigure}

\subsection{Query and Ground-truth Lineage Distribution}
\label{app:query-distribution}
Figure~\ref{fig:year-gap-by-hop} provides an additional temporal analysis for multi-hop lineage relations. Ancestor papers are on average earlier than the query paper, while successor papers are on average later. The temporal gap also grows with graph distance, indicating that longer-hop lineage targets tend to represent more distant stages of methodological development rather than arbitrary graph connections.  

\begin{figure}[h]
    \centering
    \includegraphics[width=1\linewidth]{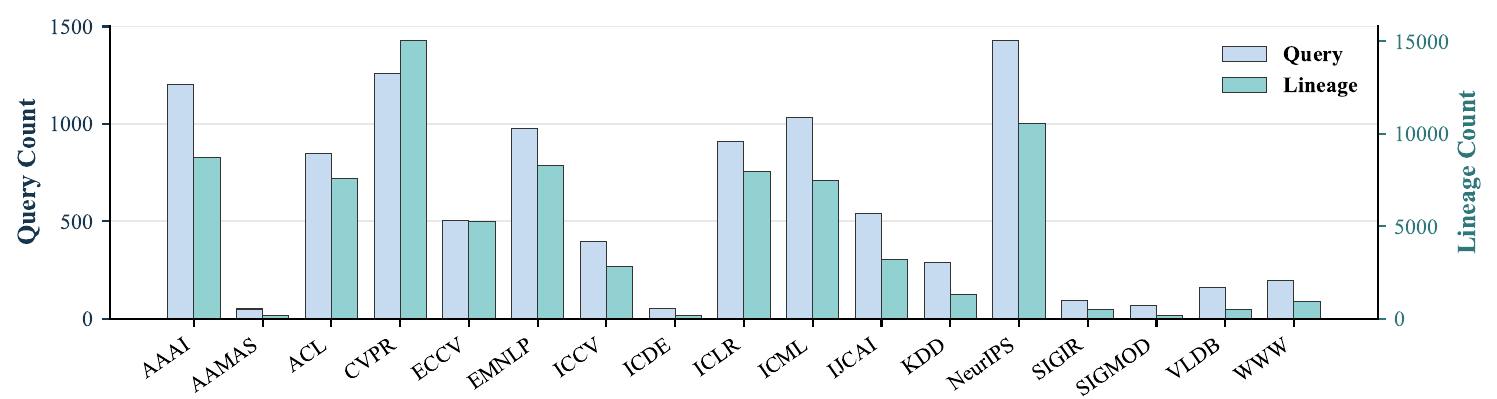}
    \caption{Venue coverage of benchmark queries and lineage ground truth.}
    \label{fig:venue-query-gt}
\end{figure}

Finally, we examine the venue distribution of benchmark queries and their lineage ground-truth papers. As shown in Figure~\ref{fig:venue-query-gt}, the query set covers all included venues and broadly follows the venue composition of the constructed corpus. The lineage ground-truth distribution is not identical to the query distribution, since each query paper may connect to prior or later papers from different venues. This cross-venue pattern reflects the fact that methodological lineage is not confined within a single publication venue and supports evaluating retrieval across the full \textsc{Lineage} corpus.

%% file: appendix/case.tex
\section{Cases}
This section provides qualitative examples of \textsc{Lineage} to clarify the benchmark semantics and usage. We illustrate how empirical baseline evidence induces lineage graphs, and how multi-facet queries express different search intents for the same query paper. 

\subsection{Lineage Graph}
\label{app:graph-case}
\begin{figure}[htb]
    \centering
    \includegraphics[width=0.87\linewidth]{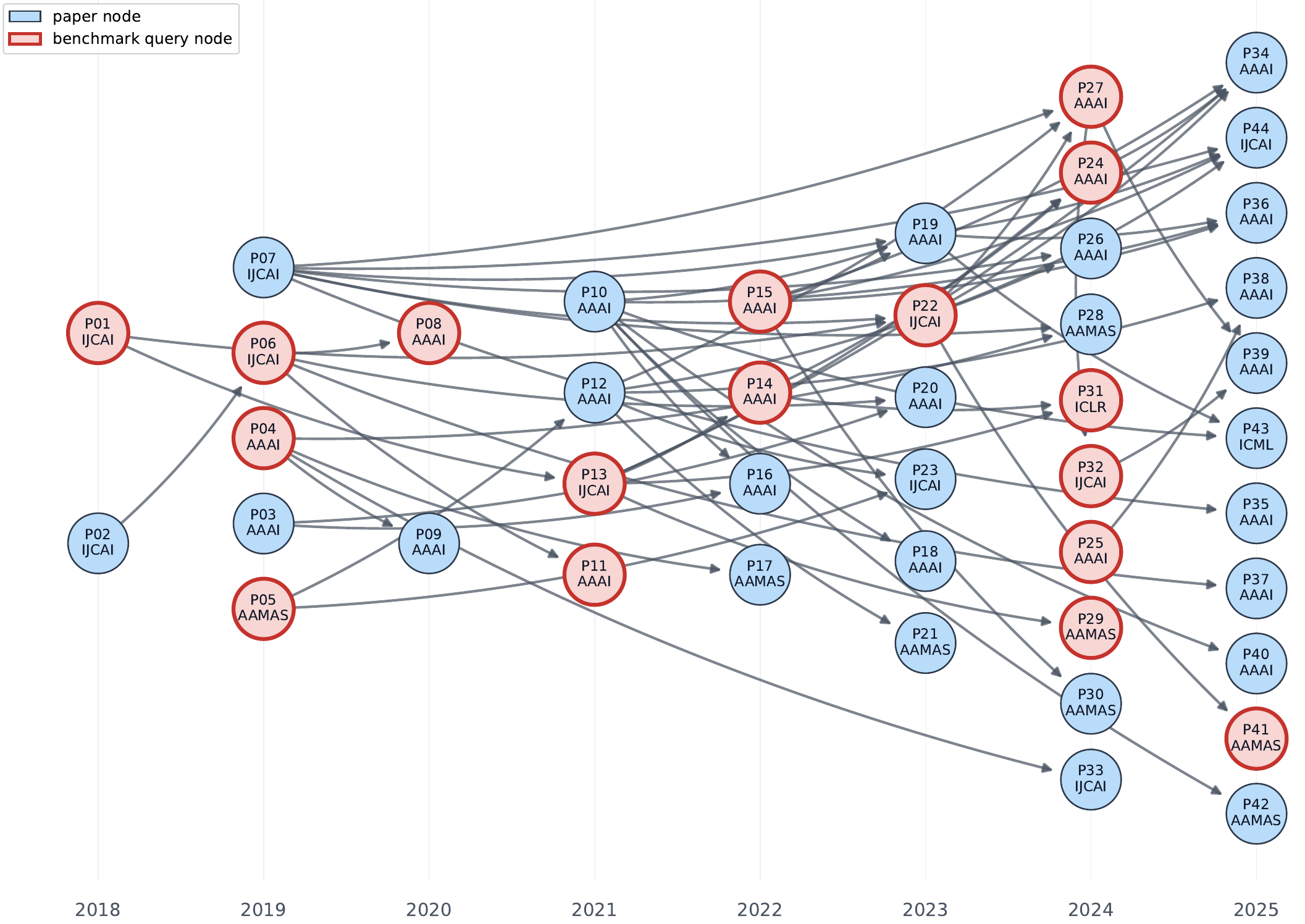}
    \caption{The subgraph of papers in \textit{Multi-Agent Path Finding}.}
    \label{fig:graph-case-0}
\end{figure}

\begin{figure}[htb]
    \centering
    \includegraphics[width=0.87\linewidth]{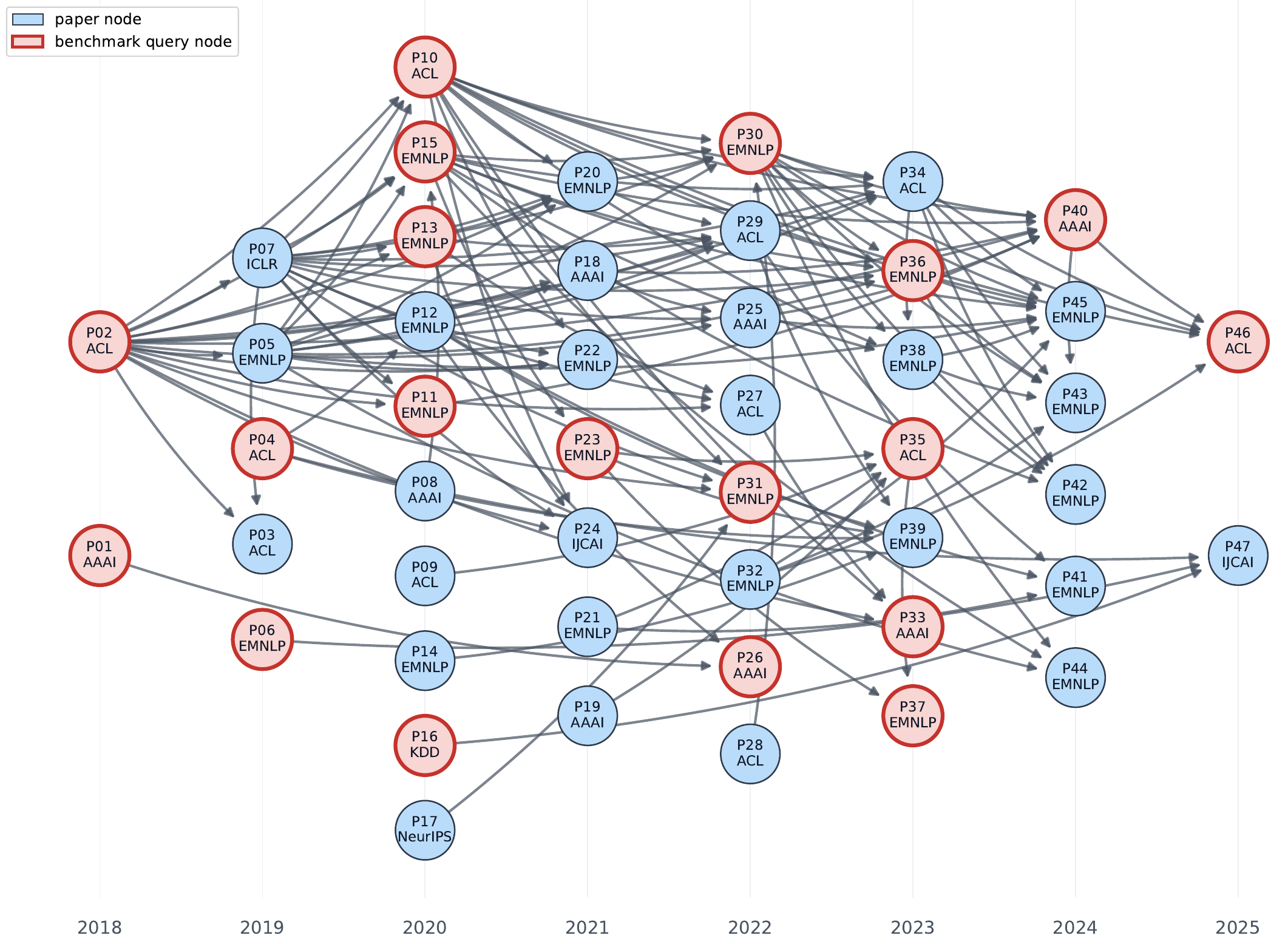}
    \caption{The subgraph of papers in \textit{Task-Oriented Dialogue Systems}.}
    \label{fig:graph-case-1}
\end{figure}

We provide two representative lineage subgraphs in Figure~\ref{fig:graph-case-0} and Figure~\ref{fig:graph-case-1} to illustrate the structure of method evolution captured by our benchmark. Each node denotes a paper, and each directed edge indicates that the paper is used as a baseline by the query paper. Benchmark query papers are highlighted in the graphs. Table~\ref{tab:case-0-list} and Table~\ref{tab:case-1-list} list the full title of each paper node.

\begin{table}[h]
\centering
\caption{Papers in the subgraph of multi-agent path finding.}
\label{tab:case-0-list}
\resizebox{\linewidth}{!}{\begin{tabular}{ll}
\toprule
\textbf{Point} & \textbf{Title} \\
\midrule
P01 & Anytime Focal Search with Applications \\
P02 & Extended Increasing Cost Tree Search for Non-Unit Cost Domains \\
P03 & Online Multi-Agent Pathfinding \\
P04 & Searching with Consistent Prioritization for Multi-Agent Path Finding \\
P05 & Task and Path Planning for Multi-Agent Pickup and Delivery \\
P06 & Multi-Agent Pathfinding with Continuous Time \\
P07 & Priority Inheritance with Backtracking for Iterative Multi-agent Path Finding \\
P08 & Generalized and Sub-Optimal Bipartite Constraints for Conflict-Based Search \\
P09 & Idle Time Optimization for Target Assignment and Path Finding in Sortation Centers \\
P10 & EECBS: A Bounded-Suboptimal Search for Multi-Agent Path Finding \\
P11 & Improving Continuous-time Conflict Based Search \\
P12 & Lifelong Multi-Agent Path Finding in Large-Scale Warehouses \\
P13 & Anytime Multi-Agent Path Finding via Large Neighborhood Search \\
P14 & Anytime Multi-Agent Path Finding via Machine Learning-Guided Large Neighborhood Search \\
P15 & MAPF-LNS2: Fast Repairing for Multi-Agent Path Finding via Large Neighborhood Search \\
P16 & Shard Systems: Scalable, Robust and Persistent Multi-Agent Path Finding with Performance Guarantees \\
P17 & Multi-Agent Path Finding for Precedence-Constrained Goal Sequences \\
P18 & Effective Integration of Weighted Cost-to-Go and Conflict Heuristic within Suboptimal CBS \\
P19 & LaCAM: Search-Based Algorithm for Quick Multi-Agent Pathfinding \\
P20 & Periodic Multi-Agent Path Planning \\
P21 & Distributed Planning with Asynchronous Execution with Local Navigation for Multi-agent Pickup and Delivery Problem \\
P22 & Improving LaCAM for Scalable Eventually Optimal Multi-Agent Pathfinding \\
P23 & Multi-Robot Coordination and Layout Design for Automated Warehousing \\
P24 & Adaptive Anytime Multi-Agent Path Finding Using Bandit-Based Large Neighborhood Search \\
P25 & Decentralized Monte Carlo Tree Search for Partially Observable Multi-Agent Pathfinding \\
P26 & Learn to Follow: Decentralized Lifelong Multi-Agent Pathfinding via Planning and Learning \\
P27 & Traffic Flow Optimisation for Lifelong Multi-Agent Path Finding \\
P28 & Attention-based Priority Learning for Limited Time Multi-Agent Path Finding \\
P29 & Confidence-Based Curriculum Learning for Multi-Agent Path Finding \\
P30 & Engineering LaCAM*: Towards Real-time, Large-scale, and Near-optimal Multi-agent Pathfinding \\
P31 & Neural Neighborhood Search for Multi-agent Path Finding \\
P32 & Guidance Graph Optimization for Lifelong Multi-Agent Path Finding \\
P33 & Scalable Mechanism Design for Multi-Agent Path Finding \\
P34 & Anytime Multi-Agent Path Finding with an Adaptive Delay-Based Heuristic \\
P35 & Concurrent Planning and Execution in Lifelong Multi-Agent Path Finding with Delay Probabilities \\
P36 & LNS2+RL: Combining Multi-agent Reinforcement Learning with Large Neighborhood Search in Multi-agent Path Finding \\
P37 & Loosely Synchronized Rule-Based Planning for Multi-Agent Path Finding with Asynchronous Actions \\
P38 & MAPF-GPT: Imitation Learning for Multi-Agent Pathfinding at Scale \\
P39 & Online Guidance Graph Optimization for Lifelong Multi-Agent Path Finding \\
P40 & Windowed MAPF with Completeness Guarantees \\
P41 & Minimizing Makespan with Conflict-Based Search for Optimal Multi-Agent Path Finding \\
P42 & Single-Agent Planning in a Multi-Agent System: A Unified Framework for Type-Based Planners \\
P43 & Hybrid Quantum-Classical Multi-Agent Pathfinding \\
P44 & Multi-Agent Corridor Generating Algorithm \\
\bottomrule
\end{tabular}}
\end{table}

\begin{table}[h]
\centering
\caption{Papers in the subgraph of task-oriented dialogue systems.}
\label{tab:case-1-list}
\resizebox{\linewidth}{!}{\begin{tabular}{ll}
\toprule
Point & Title \\
\midrule
P01 & Augmenting End-to-End Dialogue Systems With Commonsense Knowledge \\
P02 & Mem2Seq: Effectively Incorporating Knowledge Bases into End-to-End Task-Oriented Dialog Systems \\
P03 & A Working Memory Model for Task-oriented Dialog Response Generation \\
P04 & OpenDialKG: Explainable Conversational Reasoning with Attention-based Walks over Knowledge Graphs \\
P05 & Entity-Consistent End-to-end Task-Oriented Dialogue System with KB Retriever \\
P06 & Towards Knowledge-Based Recommender Dialog System \\
P07 & Global-to-local Memory Pointer Networks for Task-Oriented Dialogue \\
P08 & Task-Oriented Dialog Systems That Consider Multiple Appropriate Responses under the Same Context \\
P09 & BART: Denoising Sequence-to-Sequence Pre-training for Natural Language Generation, Translation, and Comprehension \\
P10 & Dynamic Fusion Network for Multi-Domain End-to-end Task-Oriented Dialog \\
P11 & Amalgamating Knowledge from Two Teachers for Task-oriented Dialogue System with Adversarial Training \\
P12 & AttnIO: Knowledge Graph Exploration with In-and-Out Attention Flow for Knowledge-Grounded Dialogue \\
P13 & GraphDialog: Integrating Graph Knowledge into End-to-End Task-Oriented Dialogue Systems \\
P14 & Knowledge-Grounded Dialogue Generation with Pre-trained Language Models \\
P15 & Learning Knowledge Bases with Parameters for Task-Oriented Dialogue Systems \\
P16 & Improving Conversational Recommender Systems via Knowledge Graph based Semantic Fusion \\
P17 & A Simple Language Model for Task-Oriented Dialogue \\
P18 & Exploring Auxiliary Reasoning Tasks for Task-oriented Dialog Systems with Meta Cooperative Learning \\
P19 & KG-BART: Knowledge Graph-Augmented BART for Generative Commonsense Reasoning \\
P20 & Contextualize Knowledge Bases with Transformer for End-to-end Task-Oriented Dialogue Systems \\
P21 & EARL: Informative Knowledge-Grounded Conversation Generation with Entity-Agnostic Representation Learning \\
P22 & Intention Reasoning Network for Multi-Domain End-to-end Task-Oriented Dialogue \\
P23 & Neural Path Hunter: Reducing Hallucination in Dialogue Systems via Path Grounding \\
P24 & UniMF: A Unified Framework to Incorporate Multimodal Knowledge Bases intoEnd-to-End Task-Oriented Dialogue Systems \\
P25 & GraphMemDialog: Optimizing End-to-End Task-Oriented Dialog Systems Using Graph Memory Networks \\
P26 & HiTKG: Towards Goal-Oriented Conversations via Multi-Hierarchy Learning \\
P27 & An Interpretable Neuro-Symbolic Reasoning Framework for Task-Oriented Dialogue Generation \\
P28 & Multi-Task Pre-Training for Plug-and-Play Task-Oriented Dialogue System \\
P29 & Towards Large-Scale Interpretable Knowledge Graph Reasoning for Dialogue Systems \\
P30 & Q-TOD: A Query-driven Task-oriented Dialogue System \\
P31 & Robust Task-Oriented Dialogue Generation with Contrastive Pre-training and Adversarial Filtering \\
P32 & UnifiedSKG: Unifying and Multi-Tasking Structured Knowledge Grounding with Text-to-Text Language Models \\
P33 & Towards Complex Scenarios: Building End-to-End Task-Oriented Dialogue System across Multiple Knowledge Bases \\
P34 & Multi-Grained Knowledge Retrieval for End-to-End Task-Oriented Dialog \\
P35 & RHO: Reducing Hallucination in Open-domain Dialogues with Knowledge Grounding \\
P36 & Dual-Feedback Knowledge Retrieval for Task-Oriented Dialogue Systems \\
P37 & Exploring In-Context Learning for Knowledge Grounded Dialog Generation \\
P38 & Retrieval-Generation Alignment for End-to-End Task-Oriented Dialogue System \\
P39 & Well Begun is Half Done: Generator-agnostic Knowledge Pre-Selection for Knowledge-Grounded Dialogue \\
P40 & From Retrieval to Generation: A Simple and Unified Generative Model for End-to-End Task-Oriented Dialogue \\
P41 & Generative Subgraph Retrieval for Knowledge Graph-Grounded Dialog Generation \\
P42 & Learning to Match Representations is Better for End-to-End Task-Oriented Dialog System \\
P43 & Relevance Is a Guiding Light: Relevance-aware Adaptive Learning for End-to-end Task-oriented Dialogue System \\
P44 & Synergizing In-context Learning with Hints for End-to-end Task-oriented Dialog Systems \\
P45 & What are the Generator Preferences for End-to-end Task-Oriented Dialog System? \\
P46 & LLM-Enhanced Query Generation and Retrieval Preservation for Task-Oriented Dialogue \\
P47 & Do Mentioned Items Truly Matter? Enhancing Conversational Recommender Systems with Causal Intervention and Large Language Models \\
\bottomrule
\end{tabular}}
\end{table}

\clearpage
\subsection{Multi-Facet Queries}
\label{app:query-case}
We provide two query cases (sample\_id lineage\_000234 and lineage\_001813) in Table~\ref{tab:query-example-1} and Table~\ref{tab:query-example-2} to illustrate how multi-facet queries describe the same query paper from paper-centric, topical, contextual, and methodological perspectives. 

\definecolor{queryblue}{HTML}{4AA3DF}

\begin{table*}[h]
\caption{Selected query example (ID: \textit{lineage\_000234}).}
\label{tab:query-example-1}
\centering
\footnotesize
\renewcommand{\arraystretch}{1.16}
\begin{tabular}{>{\raggedright\arraybackslash}p{0.1\textwidth}|>{\raggedright\arraybackslash}p{0.83\textwidth}}
\toprule
\textbf{Key} & \textbf{Value} \\
\midrule
\textbf{Title} & SketchGen: Generating Constrained CAD Sketches \\
\textbf{Venue} & NeurIPS \\
\textbf{Year} & 2021 \\
\textbf{Abstract} & Computer-aided design (CAD) is the most widely used modeling approach for technical design. The typical starting point in these designs is 2D sketches which can later be extruded and combined to obtain complex three-dimensional assemblies. Such sketches are typically composed of parametric primitives, such as points, lines, and circular arcs, augmented with geometric constraints linking the primitives, such as coincidence, parallelism, or orthogonality. Sketches can be represented as graphs, with the primitives as nodes and the constraints as edges. Training a model to automatically generate CAD sketches can enable several novel workflows, but is challenging due to the complexity of the graphs and the heterogeneity of the primitives and constraints ... [total 240 words] \\
\midrule
\textbf{\textit{Title-Ctx}} & SketchGen: Generating Constrained CAD Sketches. + \textcolor{queryblue}{<abstract>}. \\
\textbf{\textit{Top-KW}} & CAD sketch generation transformer. \\
\textbf{\textit{Top-Sem}} & Generative models for CAD sketch creation using transformer architectures and constraint-aware sequences. \\
\textbf{\textit{Ctx-KW}} & Parametric CAD design constraint-based modeling. \\
\textbf{\textit{Ctx-Sem}} & CAD sketch generation in technical design with parametric primitives and geometric constraints. \\
\textbf{\textit{Meth-KW}} & Transformer graph-based sketch representation. \\
\textbf{\textit{Meth-Sem}} & Transformer-based generative models for sketch graphs with constraint and primitive type encoding. \\
\bottomrule
\end{tabular}
\end{table*}

\begin{table*}[h]
\caption{Selected query example (ID: \textit{lineage\_001813}).}
\label{tab:query-example-2}
\centering
\footnotesize
\renewcommand{\arraystretch}{1.16}
\begin{tabular}{>{\raggedright\arraybackslash}p{0.1\textwidth}|>{\raggedright\arraybackslash}p{0.83\textwidth}}
\toprule
\textbf{Key} & \textbf{Value} \\
\midrule
\textbf{Title} & HUMUS-Net: Hybrid Unrolled Multi-scale Network Architecture for Accelerated MRI Reconstruction \\
\textbf{Venue} & NeurIPS \\
\textbf{Year} & 2022 \\
\textbf{Abstract} & In accelerated MRI reconstruction, the anatomy of a patient is recovered from a set of undersampled and noisy measurements. Deep learning approaches have been proven to be successful in solving this ill-posed inverse problem and are capable of producing very high quality reconstructions. However, current architectures heavily rely on convolutions, that are content-independent and have difficulties modeling long-range dependencies in images. Recently, Transformers, the workhorse of contemporary natural language processing, have emerged as powerful building blocks for a multitude of vision tasks. These models split input images into non-overlapping patches, embed the patches into lower-dimensional tokens and utilize a self-attention mechanism that does not suffer from the aforementioned weaknesses of convolutional architectures ... [total 268 words] \\
\midrule
\textbf{\textit{Title-Ctx}} & HUMUS-Net: Hybrid Unrolled Multi-scale Network Architecture for Accelerated MRI Reconstruction. + \textcolor{queryblue}{<abstract>}. \\
\textbf{\textit{Top-KW}} & MRI reconstruction unrolled network multi-scale. \\
\textbf{\textit{Top-Sem}} & Hybrid convolution-transformer networks for MRI reconstruction with multi-scale feature refinement. \\
\textbf{\textit{Ctx-KW}} & Accelerated MRI undersampled measurements noise recovery. \\
\textbf{\textit{Ctx-Sem}} & MRI reconstruction from undersampled noisy measurements in clinical imaging settings. \\
\textbf{\textit{Meth-KW}} & Hybrid convolution transformer multi-scale feature extraction. \\
\textbf{\textit{Meth-Sem}} & Unrolled multi-scale networks combining convolutional and transformer blocks for feature refinement. \\
\bottomrule
\end{tabular}
\end{table*}

%% file: appendix/prompts.tex
\newcommand{\WC}[1]{\textcolor{lightblue}{\ttfamily\detokenize{[#1]}}}
\definecolor{lightblue}{HTML}{4AA3DF}
\definecolor{titlegray}{HTML}{BFBFBF}
\definecolor{boxbg}{HTML}{F7F7F7}

\newtcolorbox{PromptStyleBox}[1]{
  enhanced,
  breakable,
  unbreakable,
  colback=boxbg,
  colframe=titlegray,
  boxrule=0.6pt,
  arc=2pt,
  left=10pt,right=10pt,top=8pt,bottom=10pt,
  title=\textbf{#1},
  fonttitle=\bfseries,
  coltitle=black,
  boxed title style={
    colback=titlegray,
    colframe=titlegray,
    arc=0pt,
    left=8pt,right=8pt,top=4pt,bottom=4pt
  }
}

\newenvironment{PromptBody}
{\parindent=0pt
  \parskip=2pt
  \raggedright
  \sloppy
  \hyphenpenalty=10000
  \exhyphenpenalty=10000
}
{\par}
\clearpage
\section{Prompts}
\label{app:prompt}
This section presents the prompts used in the LLM-based components of \textsc{Lineage}-Constructor and \textsc{Lineage} benchmark construction. These prompts cover paper type classification, dual-source lineage extraction, conflict resolution over MinerU and GROBID candidates, and multi-facet retrieval query generation. Each prompt defines the required input fields, decision rules, and JSON output format, making the automatic construction process reproducible and traceable.

\subsection{Lineage Graph Construction Prompt}
\label{app:graph-prompt}
The lineage graph construction prompts are used by the paper type classifier and the dual-source lineage extractor. The paper type classification prompt determines the primary contribution type of each paper and retains methodology papers for downstream lineage extraction. The single-source lineage extraction prompt identifies experiment-grounded lineage relations from one parsed paper source by extracting prior methods that are explicitly used as empirical comparison baselines and normalizing them to full reference paper titles. When the \texttt{MinerU} and \texttt{GROBID} candidate sets conflict, the dual-source conflict resolution prompt performs candidate-restricted integration by re-checking the candidates against the \texttt{MinerU} paper content and retaining only evidence-supported baseline titles.

\begin{figure*}[h]
\begin{PromptStyleBox}{Paper Type Classification}

\textbf{System Prompt:}

\begin{PromptBody}
You are a strict research paper type classifier. Determine the PRIMARY contribution type of the paper.

\textbf{Definitions:}
\begin{itemize}
    \item methodology: The paper proposes a novel model, algorithm, technical framework, or system. Experiments or datasets may be included, but the core contribution is a new technical method.
    \item benchmark: The paper primarily introduces a new dataset, benchmark, evaluation protocol, or leaderboard. The main contribution is evaluation infrastructure rather than a new technical method.
    \item survey: The paper systematically reviews, summarizes, or categorizes existing literature. It does NOT introduce a novel algorithm or technical framework.
    \item others: Anything that does not clearly fit the above categories.
\end{itemize}

\textbf{Priority rules:}
\begin{enumerate}
    \item If the paper proposes a novel model, algorithm, or technical framework, classify it as `methodology' even if it also includes experiments or introduces a dataset.
    \item Only classify as `benchmark' if the primary contribution is dataset or evaluation construction without a new method.
    \item Only classify as `survey' if the paper synthesizes existing work and does not propose a new method.
    \item If unsure, choose `others'.
\end{enumerate}

Respond ONLY with a JSON object of the form:\\
\{"paper\_type": "methodology|benchmark|survey|others"\}.
\end{PromptBody}

\vspace{6pt}
\textbf{User Prompt:}

\begin{PromptBody}
Title: \WC{title}

Abstract: \WC{abstract}

Introduction: \WC{introduction}
\end{PromptBody}

\end{PromptStyleBox}
\end{figure*}

\begin{figure*}[h]
\begin{PromptStyleBox}{Single-Source Lineage Extraction}

\textbf{System Prompt:}

\begin{PromptBody}
You are an expert at extracting experimental baselines from academic papers. Identify only methods that are genuinely used as comparison baselines in experiments. Exclude the proposed method, its variants, ablations, datasets, metrics, tools, and methods mentioned only in background discussion. Return only a JSON object.
\end{PromptBody}

\vspace{6pt}
\textbf{User Prompt:}

\begin{PromptBody}
\textbf{Extraction rules:}
\begin{enumerate}
    \item Scope: Extract only methods that are explicitly compared in experimental evaluation, quantitative results, qualitative comparison, or implementation/evaluation setup tied to reported comparisons.
    \item Valid baseline: A baseline must be a separate method/model/system that the paper compares against. Keep methods appearing in comparison tables, result figures, experimental setup, or evaluation text.
    \item Exclude: the proposed method, ablation variants, internal modules, training tricks, datasets, metrics, loss terms, tools, and related-work citations that are not actually evaluated.
    \item References are supporting evidence. The output must be the full paper title of the cited baseline work from the References section, not a method name, model nickname, abbreviation, or benchmark label.
    \item If several baseline method names or variants in the paper map to the same cited paper, output that paper title only once.
    \item Use references only to map evidence-supported compared methods to full paper titles. Do not output a paper title unless some provided evidence unit directly supports that method as an evaluated comparison baseline.
    \item Method abbreviations, aliases, or shorthand names in evidence units may correspond to full paper titles in the references. Resolve them conservatively.
    \item Do not output papers mentioned only as background, inspiration, task framing, benchmark context, or general related work.
    \item Do not output generic backbone, toolkit, dataset, or optimization papers unless the evidence units explicitly present them as compared baselines.
    \item Some papers are theoretical papers or do not use standard baseline comparisons. If no valid baselines are found, return an empty list.
    \item Output only JSON in the format \{"baseline\_titles": ["Full Reference Paper Title A", "Full Reference Paper Title B"]\}. Return \{"baseline\_titles": []\} if none are found.
\end{enumerate}

\textbf{Paper content:}

Source mode: \WC{source_mode}

\WC{paper_content}

Return only a JSON object with key baseline\_titles.
\end{PromptBody}

\end{PromptStyleBox}
\end{figure*}

\begin{figure*}[h]
\begin{PromptStyleBox}{Dual-Source Conflict Resolution}

\textbf{System Prompt:}

\begin{PromptBody}
You are an expert at extracting experimental baselines from academic papers. Identify only methods that are genuinely used as comparison baselines in experiments. Exclude the proposed method, its variants, ablations, datasets, metrics, tools, and methods mentioned only in background discussion. Return only a JSON object.
\end{PromptBody}

\vspace{6pt}
\textbf{User Prompt:}

\begin{PromptBody}
\textbf{Extraction rules:}

\WC{extraction_rules}

The baseline lists extracted from two parser sources conflict. Re-check the MinerU paper content and keep only baseline paper titles that are truly supported. Prefer a supported subset from the provided candidate titles. Do not invent new titles.

\textbf{MinerU paper content:}

\WC{mineru_paper_content}

\textbf{MinerU candidate titles:}

\WC{mineru_candidate_titles}

\textbf{GROBID candidate titles:}

\WC{grobid_candidate_titles}

Return only JSON: \{"baseline\_titles": [str, ...]\}.
\end{PromptBody}

\end{PromptStyleBox}
\end{figure*}

\subsection{Query Generation Prompt}
\label{app:query-prompt}
The query generation prompt is used to construct multi-facet retrieval queries for \textsc{Lineage}. Given a paper title and abstract, the prompt generates short academic retrieval queries that describe the paper's research problem, task setting or application context, and method family or technical mechanism. These generated queries instantiate the topical, contextual, and methodological facets, while the title and title-context queries are directly derived from the original metadata.

\begin{figure*}[h]
\begin{PromptStyleBox}{Query Generation}

\textbf{System Prompt:}

\begin{PromptBody}
You generate short academic retrieval queries from paper title and abstract. The query should be used to retrieve related papers in the same research area, not the source paper itself.

\WC{style_instruction}

\textbf{Generate one query for each topic below:}
\begin{itemize}
    \item abstract: Describe the paper's research problem and broad technical direction.
    \item background: Describe the task setting, motivation, or application context.
    \item methodology: Describe the method family, modeling idea, or technical mechanism.
\end{itemize}

\textbf{Requirements:}
\begin{itemize}
    \item Keep each query short.
    \item Keep 1 to 2 concrete domain anchors, chosen from problem setting, method family, or technical mechanism.
    \item Use shared research terms instead of title-like, paper-specific, self-referential, or filler wording; if a method name is too specific, replace it with its broader method family or mechanism, while preserving the main technical topic.
    \item Background queries should contain the problem context; methodology queries should contain the method family or core technical mechanism.
    \item Do not add facts not present in the given title or abstract.
    \item If unsure, make the query slightly broader rather than narrower.
    \item Return valid JSON only.
    \item Use exactly these keys: \WC{expected_fields}.
\end{itemize}
\end{PromptBody}

\vspace{6pt}
\textbf{User Prompt:}

\begin{PromptBody}
Title: \WC{title}

Abstract: \WC{abstract}
\end{PromptBody}

\end{PromptStyleBox}
\end{figure*}

\subsection{LLM Search Prompt}
\label{app:llm-search-prompt}
The LLM search prompt is used to evaluate LLM-based web search baselines on \textsc{Lineage}. Given the query paper metadata and a retrieval query, the model is asked to search for real methodology papers that match the query while excluding the query paper itself. The prompt prioritizes papers that propose concrete models, algorithms, methods, or technical frameworks, while discouraging surveys, benchmarks, datasets, and purely analytical papers. It constrains returned papers to the allowed venues and the 2018--2025 publication range, and requires the model to output only a ranked JSON list of paper titles.

\begin{figure*}[h]
\begin{PromptStyleBox}{LLM Search Prompt}

\textbf{System Prompt:}

\begin{PromptBody}
You are an academic paper search assistant focused on methodology papers.
Use a web search to find real methodology papers matching the search query. Return only JSON.

Prioritize papers that propose a concrete model, algorithm, method, or technical framework.
Avoid surveys, benchmarks, datasets, and purely analytical papers unless they introduce a clear new method.
The result list must be ranked from most relevant to least relevant.
Every returned paper must be published in one of the allowed venues and between 2018 and 2025 inclusive.
Do not include the query paper itself.
Return paper titles only; do not return venue, year, URL, reason, abstract, authors, or any other field.
Do not include any extra text before or after the JSON object. Do not use the thinking mode.
\end{PromptBody}

\vspace{6pt}
\textbf{User Prompt:}

\begin{PromptBody}
Query paper:
\WC{query_paper}

Search query:
\WC{search_query}

Allowed venues:
\WC{allowed_venues}

Return at least \WC{top_k} ranked paper titles unless fewer than \WC{top_k} valid matching papers exist.
Return JSON with this exact shape:
\{
  "titles": [
    "most relevant paper title",
    "next most relevant paper title"
  ]
\}
Do not include any other text, markdown, explanation, code fences, or additional keys.
\end{PromptBody}

\end{PromptStyleBox}
\end{figure*}

%% file: ref.bib
@inproceedings{chenbrowsecomp,
  title={BrowseComp-Plus: A More Fair and Transparent Evaluation Benchmark of Deep-Research Agent},
  author={Chen, Zijian and Ma, Xueguang and Zhuang, Shengyao and Nie, Ping and Zou, Kai and Sharifymoghaddam, Sahel and Liu, Andrew and Green, Joshua and Patel, Kshama and Meng, Ruoxi and others},
  booktitle={First Workshop on Multi-Turn Interactions in Large Language Models},
  year={2025}
}

@inproceedings{tangai,
  title={AI-Researcher: Autonomous Scientific Innovation},
  author={Tang, Jiabin and Xia, Lianghao and Li, Zhonghang and Huang, Chao},
  booktitle={The Thirty-ninth Annual Conference on Neural Information Processing Systems},
  year={2025}
}

@inproceedings{yang2026multimodal,
  title={Multimodal deepresearcher: Generating text-chart interleaved reports from scratch with agentic framework},
  author={Yang, Zhaorui and Pan, Bo and Wang, Han and Wang, Yiyao and Liu, Xingyu and Weng, Luoxuan and Feng, Yingchaojie and Feng, Haozhe and Zhu, Minfeng and Zhang, Bo and others},
  booktitle={Proceedings of the AAAI Conference on Artificial Intelligence},
  volume={40},

  pages={34368--34377},
  year={2026}
}

@inproceedings{he2025pasa,
  title={Pasa: An llm agent for comprehensive academic paper search},
  author={He, Yichen and Huang, Guanhua and Feng, Peiyuan and Lin, Yuan and Zhang, Yuchen and Li, Hang and others},
  booktitle={Proceedings of the 63rd Annual Meeting of the Association for Computational Linguistics (Volume 1: Long Papers)},
  pages={11663--11679},
  year={2025}
}

@article{lu2026towards,
  title={Towards end-to-end automation of AI research},
  author={Lu, Chris and Lu, Cong and Lange, Robert Tjarko and Yamada, Yutaro and Hu, Shengran and Foerster, Jakob and Ha, David and Clune, Jeff},
  journal={Nature},
  volume={651},
  number={8107},
  pages={914--919},
  year={2026},
  publisher={Nature Publishing Group UK London}
}

@inproceedings{
zhu2026autofigure,
title={AutoFigure: Generating and Refining Publication-Ready Scientific Illustrations},
author={Minjun Zhu and Zhen Lin and Yixuan Weng and Panzhong Lu and Qiujie Xie and Yifan Wei and Sifan Liu and QiYao Sun and Yue Zhang},
booktitle={The Fourteenth International Conference on Learning Representations},
year={2026}
}

@inproceedings{chenmlr,
  title={MLR-Bench: Evaluating AI Agents on Open-Ended Machine Learning Research},
  author={Chen, Hui and Xiong, Miao and Lu, Yujie and Han, Wei and Deng, Ailin and He, Yufei and Wu, Jiaying and Li, Yibo and Liu, Yue and Hooi, Bryan},
  booktitle={The Thirty-ninth Annual Conference on Neural Information Processing Systems Datasets and Benchmarks Track},
  year={2025}
}

@article{pan2026paperscout,
  title={Paperscout: An autonomous agent for academic paper search with process-aware sequence-level policy optimization},
  author={Pan, Tingyue and Ouyang, Jie and Cheng, Mingyue and Li, Qingchuan and Liu, Zirui and Wang, Daoyu and Pan, Mingfan and Yu, Shuo and Liu, Qi},
  journal={arXiv preprint arXiv:2601.10029},
  year={2026}
}

@article{chen2026mars,
  title={MARS: Modular Agent with Reflective Search for Automated AI Research},
  author={Chen, Jiefeng and Mishra, Bhavana Dalvi and Nam, Jaehyun and Meng, Rui and Pfister, Tomas and Yoon, Jinsung},
  journal={arXiv preprint arXiv:2602.02660},
  year={2026}
}

@article{majumder2024discoverybench,
  title={Discoverybench: Towards data-driven discovery with large language models},
  author={Majumder, Bodhisattwa Prasad and Surana, Harshit and Agarwal, Dhruv and Mishra, Bhavana Dalvi and Meena, Abhijeetsingh and Prakhar, Aryan and Vora, Tirth and Khot, Tushar and Sabharwal, Ashish and Clark, Peter},
  journal={arXiv preprint arXiv:2407.01725},
  year={2024}
}

@inproceedings{chenscienceagentbench,
  title={ScienceAgentBench: Toward Rigorous Assessment of Language Agents for Data-Driven Scientific Discovery},
  author={Chen, Ziru and Chen, Shijie and Ning, Yuting and Zhang, Qianheng and Wang, Boshi and Yu, Botao and Li, Yifei and Liao, Zeyi and Wei, Chen and Lu, Zitong and others},
  booktitle={The Thirteenth International Conference on Learning Representations},
  year = {2025}
}

@inproceedings{huang2024mlagentbench,
  title={MLAgentBench: Evaluating Language Agents on Machine Learning Experimentation},
  author={Huang, Qian and Vora, Jian and Liang, Percy and Leskovec, Jure},
  booktitle={International Conference on Machine Learning},
  pages={20271--20309},
  year={2024},
  organization={PMLR}
}

@article{asai2026synthesizing,
  title={Synthesizing scientific literature with retrieval-augmented language models},
  author={Asai, Akari and He, Jacqueline and Shao, Rulin and Shi, Weijia and Singh, Amanpreet and Chang, Joseph Chee and Lo, Kyle and Soldaini, Luca and Feldman, Sergey and D’Arcy, Mike and others},
  journal={Nature},
  pages={1--7},
  year={2026},
  publisher={Nature Publishing Group UK London}
}

@inproceedings{lo2020s2orc,
  title={S2ORC: The semantic scholar open research corpus},
  author={Lo, Kyle and Wang, Lucy Lu and Neumann, Mark and Kinney, Rodney and Weld, Daniel S},
  booktitle={Proceedings of the 58th annual meeting of the association for computational linguistics},
  pages={4969--4983},
  year={2020}
}

@inproceedings{cohan2020specter,
  title={Specter: Document-level representation learning using citation-informed transformers},
  author={Cohan, Arman and Feldman, Sergey and Beltagy, Iz and Downey, Doug and Weld, Daniel S},
  booktitle={Proceedings of the 58th annual meeting of the association for computational linguistics},
  pages={2270--2282},
  year={2020}
}

@inproceedings{singh2023scirepeval,
  title={Scirepeval: A multi-format benchmark for scientific document representations},
  author={Singh, Amanpreet and D’Arcy, Mike and Cohan, Arman and Downey, Doug and Feldman, Sergey},
  booktitle={Proceedings of the 2023 Conference on Empirical Methods in Natural Language Processing},
  pages={5548--5566},
  year={2023}
}

@article{wang2023scientific,
  title={Scientific document retrieval using multi-level aspect-based queries},
  author={Wang, Jianyou Andre and Wang, Kaicheng and Wang, Xiaoyue and Naidu, Prudhviraj and Bergen, Leon and Paturi, Ramamohan},
  journal={Advances in Neural Information Processing Systems},
  volume={36},
  pages={38404--38419},
  year={2023}
}

@inproceedings{ajith2024litsearch,
  title={Litsearch: A retrieval benchmark for scientific literature search},
  author={Ajith, Anirudh and Xia, Mengzhou and Chevalier, Alexis and Goyal, Tanya and Chen, Danqi and Gao, Tianyu},
  booktitle={Proceedings of the 2024 Conference on Empirical Methods in Natural Language Processing},
  pages={15068--15083},
  year={2024}
}

@article{wu2024stark,
  title={Stark: Benchmarking llm retrieval on textual and relational knowledge bases},
  author={Wu, Shirley and Zhao, Shiyu and Yasunaga, Michihiro and Huang, Kexin and Cao, Kaidi and Huang, Qian and Ioannidis, Vassilis N and Subbian, Karthik and Zou, James and Leskovec, Jure},
  journal={Advances in Neural Information Processing Systems},
  volume={37},
  pages={127129--127153},
  year={2024}
}

@article{shao2025scinetbench,
  title={SciNetBench: A Relation-Aware Benchmark for Scientific Literature Retrieval Agents},
  author={Shao, Chenyang and Li, Yong and Xu, Fengli},
  journal={arXiv preprint arXiv:2601.03260},
  year={2025}
}

@inproceedings{kang2018dataset,
  title={A dataset of peer reviews (PeerRead): Collection, insights and NLP applications},
  author={Kang, Dongyeop and Ammar, Waleed and Dalvi, Bhavana and Van Zuylen, Madeleine and Kohlmeier, Sebastian and Hovy, Eduard and Schwartz, Roy},
  booktitle={Proceedings of the 2018 conference of the North American chapter of the Association for Computational Linguistics: Human language technologies, volume 1 (long papers)},
  pages={1647--1661},
  year={2018}
}

@inproceedings{long2024recommending,
  title={Recommending Missed Citations Identified by Reviewers: A New Task, Dataset and Baselines},
  author={Long, Kehan and Li, Shasha and Wang, Pancheng and Bao, Chenlong and Tang, Jintao and Wang, Ting},
  booktitle={Proceedings of the 2024 Joint International Conference on Computational Linguistics, Language Resources and Evaluation (LREC-COLING 2024)},
  pages={13699--13711},
  year={2024}
}

@article{press2024citeme,
  title={CiteME: Can Language Models Accurately Cite Scientific Claims?},
  author={Press, Ori and Hochlehnert, Andreas and Prabhu, Ameya and Udandarao, Vishaal and Press, Ofir and Bethge, Matthias},
  journal={Advances in Neural Information Processing Systems},
  volume={37},
  pages={7847--7877},
  year={2024}
}

@inproceedings{zheng2026should,
  title={What should i cite? a rag benchmark for academic citation prediction},
  author={Zheng, Leqi and Zhang, Jiajun and Chen, Canzhi and Wang, Chaokun and Li, Hongwei and Li, Yuying and Mao, Yaoxin and Yan, Shannan and Song, Zixin and Feng, Zhiyuan and others},
  booktitle={Proceedings of the ACM Web Conference 2026},
  pages={1852--1863},
  year={2026}
}

@inproceedings{wadden2020fact,
  title={Fact or fiction: Verifying scientific claims},
  author={Wadden, David and Lin, Shanchuan and Lo, Kyle and Wang, Lucy Lu and van Zuylen, Madeleine and Cohan, Arman and Hajishirzi, Hannaneh},
  booktitle={Proceedings of the 2020 Conference on Empirical Methods in Natural Language Processing (EMNLP)},
  pages={7534--7550},
  year={2020}
}

@inproceedings{dasigi2021dataset,
  title={A dataset of information-seeking questions and answers anchored in research papers},
  author={Dasigi, Pradeep and Lo, Kyle and Beltagy, Iz and Cohan, Arman and Smith, Noah A and Gardner, Matt},
  booktitle={Proceedings of the 2021 Conference of the North American Chapter of the Association for Computational Linguistics: Human Language Technologies},
  pages={4599--4610},
  year={2021}
}

@inproceedings{lee2023qasa,
  title={Qasa: advanced question answering on scientific articles},
  author={Lee, Yoonjoo and Lee, Kyungjae and Park, Sunghyun and Hwang, Dasol and Kim, Jaehyeon and Lee, Hong-in and Lee, Moontae},
  booktitle={International Conference on Machine Learning},
  pages={19036--19052},
  year={2023},
  organization={PMLR}
}

@inproceedings{li2024m3sciqa,
  title={M3SciQA: A multi-modal multi-document scientific QA benchmark for evaluating foundation models},
  author={Li, Chuhan and Shangguan, Ziyao and Zhao, Yilun and Li, Deyuan and Liu, Yixin and Cohan, Arman},
  booktitle={Findings of the Association for Computational Linguistics: EMNLP 2024},
  pages={15419--15446},
  year={2024}
}

@article{pramanick2024spiqa,
  title={Spiqa: A dataset for multimodal question answering on scientific papers},
  author={Pramanick, Shraman and Chellappa, Rama and Venugopalan, Subhashini},
  journal={Advances in Neural Information Processing Systems},
  volume={37},
  pages={118807--118833},
  year={2024}
}

@inproceedings{sican,
  title={Can LLMs Generate Novel Research Ideas? A Large-Scale Human Study with 100+ NLP Researchers},
  author={Si, Chenglei and Yang, Diyi and Hashimoto, Tatsunori},
  booktitle={The Thirteenth International Conference on Learning Representations},
  year={2025}
}

@inproceedings{toledoai,
  title={AI Research Agents for Machine Learning: Search, Exploration, and Generalization in MLE-bench},
  author={Toledo, Edan and Hambardzumyan, Karen and Josifoski, Martin and HAZRA, RISHI and Baldwin, Nicolas and Audran-Reiss, Alexis and Kuchnik, Michael and Magka, Despoina and Jiang, Minqi and Lupidi, Alisia Maria and others},
  booktitle={The Thirty-ninth Annual Conference on Neural Information Processing Systems},
  year={2025}
}

@inproceedings{wengcycleresearcher,
  title={CycleResearcher: Improving Automated Research via Automated Review},
  author={Weng, Yixuan and Zhu, Minjun and Bao, Guangsheng and Zhang, Hongbo and Wang, Jindong and Zhang, Yue and Yang, Linyi},
  booktitle={The Thirteenth International Conference on Learning Representations},
  year={2025}
}

@article{wang2024autosurvey,
  title={Autosurvey: Large language models can automatically write surveys},
  author={Wang, Yidong and Guo, Qi and Yao, Wenjin and Zhang, Hongbo and Zhang, Xin and Wu, Zhen and Zhang, Meishan and Dai, Xinyu and Zhang, Min and Wen, Qingsong and others},
  journal={Advances in neural information processing systems},
  volume={37},
  pages={115119--115145},
  year={2024}
}

@inproceedings{jiang2025hibench,
  title={Hibench: Benchmarking llms capability on hierarchical structure reasoning},
  author={Jiang, Zhuohang and Wu, Pangjing and Liang, Ziran and Chen, Peter Q and Yuan, Xu and Jia, Ye and Tu, Jiancheng and Li, Chen and Ng, Peter HF and Li, Qing},
  booktitle={Proceedings of the 31st ACM SIGKDD Conference on Knowledge Discovery and Data Mining V. 2},
  pages={5505--5515},
  year={2025}
}

@inproceedings{yuan2026mkg,
  title={mKG-RAG: Leveraging Multimodal Knowledge Graphs in Retrieval-Augmented Generation for Knowledge-intensive VQA},
  author={Yuan, Xu and Ning, Liangbo and Ye, Qingqing and Fan, Wenqi and Li, Qing},
  booktitle={Proceedings of the 49th International ACM SIGIR Conference on Research and Development in Information Retrieval},
  pages={2274--2285},
  year={2026}
}

@article{song2026paperorchestra,
  title={PaperOrchestra: A Multi-Agent Framework for Automated AI Research Paper Writing},
  author={Song, Yiwen and Song, Yale and Pfister, Tomas and Yoon, Jinsung},
  journal={arXiv preprint arXiv:2604.05018},
  year={2026}
}

@article{lyu2026evoscientist,
  title={Evoscientist: Towards multi-agent evolving ai scientists for end-to-end scientific discovery},
  author={Lyu, Yougang and Zhang, Xi and Yi, Xinhao and Zhao, Yuyue and Guo, Shuyu and Hu, Wenxiang and Piotrowski, Jan and Kaliski, Jakub and Urbani, Jacopo and Meng, Zaiqiao and others},
  journal={arXiv preprint arXiv:2603.08127},
  year={2026}
}

@article{xu2026train,
  title={How to Train Your Deep Research Agent? Prompt, Reward, and Policy Optimization in Search-R1},
  author={Xu, Yinuo and Lu, Shuo and Cheng, Jianjie and Wang, Meng and Xie, Qianlong and Wang, Xingxing and He, Ran and Liang, Jian},
  journal={arXiv preprint arXiv:2602.19526},
  year={2026}
}

@article{blondel2008fast,
  title={Fast unfolding of communities in large networks},
  author={Blondel, Vincent D and Guillaume, Jean-Loup and Lambiotte, Renaud and Lefebvre, Etienne},
  journal={Journal of statistical mechanics: theory and experiment},
  volume={2008},
  number={10},
  pages={P10008},
  year={2008}
}

@article{tarjan1974note,
  title={A note on finding the bridges of a graph},
  author={Tarjan, R Endre},
  journal={Information Processing Letters},
  volume={2},
  number={6},
  pages={160--161},
  year={1974},
  publisher={Elsevier}
}

@article{chatfield1986exploratory,
  title={Exploratory data analysis},
  author={Chatfield, Chris},
  journal={European journal of operational research},
  volume={23},
  number={1},
  pages={5--13},
  year={1986},
  publisher={Elsevier}
}

@article{robertson2009probabilistic,
  title={The Probabilistic Relevance Framework: BM25 and Beyond},
  author={Robertson, Stephen and Zaragoza, Hugo},
  journal={Information Retrieval},
  volume={3},
  number={4},
  pages={333--389},
  year={2009}
}

@article{izacard2022unsupervised,
title={Unsupervised Dense Information Retrieval with Contrastive Learning},
author={Gautier Izacard and Mathilde Caron and Lucas Hosseini and Sebastian Riedel and Piotr Bojanowski and Armand Joulin and Edouard Grave},
journal={Transactions on Machine Learning Research},
issn={2835-8856},
year={2022}
}

@inproceedings{chen2024m3,
  title={M3-embedding: Multi-linguality, multi-functionality, multi-granularity text embeddings through self-knowledge distillation},
  author={Chen, Jianlyu and Xiao, Shitao and Zhang, Peitian and Luo, Kun and Lian, Defu and Liu, Zheng},
  booktitle={Findings of the association for computational linguistics: ACL 2024},
  pages={2318--2335},
  year={2024}
}

@inproceedings{lee2025nvembed,
    title={{NV}-Embed: Improved Techniques for Training {LLM}s as Generalist Embedding Models},
    author={Chankyu Lee and Rajarshi Roy and Mengyao Xu and Jonathan Raiman and Mohammad Shoeybi and Bryan Catanzaro and Wei Ping},
    booktitle={The Thirteenth International Conference on Learning Representations},
    year={2025}
}

@article{akram2026jina,
  title={jina-embeddings-v5-text: Task-Targeted Embedding Distillation},
  author={Akram, Mohammad Kalim and Sturua, Saba and Havriushenko, Nastia and Herreros, Quentin and G{\"u}nther, Michael and Werk, Maximilian and Xiao, Han},
  journal={arXiv preprint arXiv:2602.15547},
  year={2026}
}

@article{zhang2025qwen3,
  title={Qwen3 embedding: Advancing text embedding and reranking through foundation models},
  author={Zhang, Yanzhao and Li, Mingxin and Long, Dingkun and Zhang, Xin and Lin, Huan and Yang, Baosong and Xie, Pengjun and Yang, An and Liu, Dayiheng and Lin, Junyang and others},
  journal={arXiv preprint arXiv:2506.05176},
  year={2025}
}

@article{sparck1972statistical,
  title={A statistical interpretation of term specificity and its application in retrieval},
  author={Sparck Jones, Karen},
  journal={Journal of documentation},
  volume={28},
  number={1},
  pages={11--21},
  year={1972},
  publisher={MCB UP Ltd}
}

@misc{dblp,
  key={DBLP},
  title        = {{DBLP Computer Science Bibliography}},
  howpublished = {\url{https://dblp.org}},
  note         = {2026}
}

@misc{openreview,
  key={OpenReview},
  title        = {{OpenReview}},
  howpublished = {\url{https://openreview.net}},
  note         = {2026}
}

@misc{arxiv,
  key={arxiv},
  title        = {{arXiv}},
  howpublished = {\url{https://arxiv.org}},
  note         = {2026}
}

@inproceedings{shao2025reasonir,
    title={Reason{IR}: Training Retrievers for Reasoning Tasks},
    author={Rulin Shao and Rui Qiao and Varsha Kishore and Niklas Muennighoff and Xi Victoria Lin and Daniela Rus and Bryan Kian Hsiang Low and Sewon Min and Wen-tau Yih and Pang Wei Koh and Luke Zettlemoyer},
    booktitle={Second Conference on Language Modeling},
    year={2025}
}

@article{team2025tongyi,
  title={Tongyi deepresearch technical report},
  author={Team, Tongyi DeepResearch and Li, Baixuan and Zhang, Bo and Zhang, Dingchu and Huang, Fei and Li, Guangyu and Chen, Guoxin and Yin, Huifeng and Wu, Jialong and Zhou, Jingren and others},
  journal={arXiv preprint arXiv:2510.24701},
  year={2025}
}

@article{chen2026agentir,
  title={AgentIR: Reasoning-Aware Retrieval for Deep Research Agents},
  author={Chen, Zijian and Ma, Xueguang and Zhuang, Shengyao and Lin, Jimmy and Asai, Akari and Zhong, Victor},
  journal={arXiv preprint arXiv:2603.04384},
  year={2026}
}

@article{wang2024mineru,
  title={Mineru: An open-source solution for precise document content extraction},
  author={Wang, Bin and Xu, Chao and Zhao, Xiaomeng and Ouyang, Linke and Wu, Fan and Zhao, Zhiyuan and Xu, Rui and Liu, Kaiwen and Qu, Yuan and Shang, Fukai and others},
  journal={arXiv preprint arXiv:2409.18839},
  year={2024}
}

@misc{grobid,
    title = {GROBID},
    howpublished = {\url{https://github.com/grobidOrg/grobid}},
    publisher = {GitHub},
    date = {2008/},
    archivePrefix = {swh},
    eprint = {1:dir:dab86b296e3c3216e2241968f0d63b68e8209d3c}
}

@article{watts1998collective,
  title={Collective dynamics of ‘small-world’networks},
  author={Watts, Duncan J and Strogatz, Steven H},
  journal={nature},
  volume={393},
  number={6684},
  pages={440--442},
  year={1998},
  publisher={Nature Publishing Group}
}

@article{onnela2007structure,
  title={Structure and tie strengths in mobile communication networks},
  author={Onnela, J-P and Saram{\"a}ki, Jari and Hyv{\"o}nen, Jorkki and Szab{\'o}, Gy{\"o}rgy and Lazer, David and Kaski, Kimmo and Kert{\'e}sz, J{\'a}nos and Barab{\'a}si, A-L},
  journal={Proceedings of the national academy of sciences},
  volume={104},
  number={18},
  pages={7332--7336},
  year={2007},
  publisher={National Academy of Sciences}
}

@article{yuan2026magic,
  title={MAGIC: Marginal-Guided Compression with Optimal Transport for Efficient Visual Document Retrieval},
  author={Yuan, Xu and Liu, Hua and Fan, Wenqi and Li, Qing},
  journal={arXiv preprint arXiv:2609.21018},
  year={2026}
}

@article{jiang2025qa,
  title={Qa-dragon: Query-aware dynamic rag system for knowledge-intensive visual question answering},
  author={Jiang, Zhuohang and Wu, Pangjing and Yuan, Xu and Fan, Wenqi and Li, Qing},
  journal={arXiv preprint arXiv:2508.05197},
  year={2025}
}

@inproceedings{wu2026datamart,
  title={Datamart-Agent: LLM-Driven Game-Theoretic Agent for Data Marketplace Modeling},
  author={Wu, Pangjing and Chen, Peter Q and Li, Xiaodong and Fan, Wenqi and Li, Qing},
  booktitle={Findings of the Association for Computational Linguistics: ACL 2026},
  pages={32509--32531},
  year={2026}
}

@article{wu2026llm,
  title={LLM-Powered KG Utility Evaluation Framework},
  author={Wu, Pangjing and Li, Xiaodong and Fan, Wenqi and Li, Qing},
  journal={IEEE Transactions on Knowledge and Data Engineering},
  year={2026},
  publisher={IEEE},
  pages={1--14}
}
